\documentclass[12pt]{article}
\usepackage{makeidx}
\usepackage{amssymb}
\usepackage{amsfonts}
\usepackage{amsmath}
\usepackage{graphicx}
\usepackage{setspace}
\usepackage{authblk}
\usepackage{units}
\usepackage{appendix}
\usepackage[left=2.5cm,top=2.5cm,right=2.5cm,bottom=2cm]{geometry}
\usepackage{xcolor}
\usepackage[hyperfootnotes=false]{hyperref}
\usepackage{soul}
\usepackage{revsymb} 
\hypersetup{colorlinks=true,linkcolor=black,citecolor=black,urlcolor=black}

\begin{document}

\title{{\textbf{Semiclassical radiation spectrum from an electron in an external plane wave field}}}
\author[1]{T. C. Adorno\thanks{Tiago.Adorno@xjtlu.edu.cn}}
\author[2]{A. J. D. Farias Junior\thanks{antonio.farias@ifal.edu.br}}
\author[3,4]{D. M. Gitman\thanks{dmitrygitman@hotmail.com}}
\affil[1]{\textit{Department of Physics, School of Mathematics and Physics, Xi'an Jiaotong-Liverpool University, 111 Ren'ai Road, Suzhou Dushu Lake Science and Education Innovation District, Suzhou Industrial Park, Suzhou 215123, People's Republic of China;}}
\affil[2]{\textit{Instituto Federal de
Alagoas, CEP: 57466-034, Piranhas, Alagoas, Brazil;}}
\affil[3]{\textit{P. N. Lebedev Physical Institute, 53 Leninskiy prospekt, 119991, Moscow, Russia;}}
\affil[4]{\textit{Instituto de F\'isica, Universidade de S\~{a}o Paulo, Caixa Postal 66318, CEP 05508-090, S\~{a}o
Paulo, S.P., Brazil.}}

\maketitle

\onehalfspacing

\begin{abstract}
In this work, we study the electromagnetic energy and energy rate spectra
produced by a point particle in the presence of plane wave fields. Our
approach is based on a semiclassical formulation, in which the current distribution that generates electromagnetic radiation is treated classically while the radiation field is quantum. Unlike the classical energy spectrum--which exhibits divergences linked to the duration of interaction between the particle and the external field--the semiclassical spectrum is finite because radiation is produced during the quantum transition from an initial state without photons to the final state with photons at time $t$. In our formulation, we find that the maximum energy spectrum emitted by the particle is linearly proportional to time or phase, depending on the external field. This allowed us not only to extract the maximum energy rate spectra emitted by the particle but also to correlate them with energy rates derived in the framework of Classical Electrodynamics and Quantum Electrodynamics.
\end{abstract}

\section{Introduction\label{Sec1}}

Electromagnetic radiation emitted by charged distributions under the
influence of external sources, such as background fields, is a fundamental
phenomenon that occurs in a wide range of natural and technological
processes. Synchrotron radiation, for instance, is commonly observed in
astrophysical environments. It occurs when charged particles are rapidly
accelerated by strong magnetic fields, such as during intense gamma-ray
emissions from Blazars \cite{Aharonian00}, from energy dissipation events by
Pulsars \cite{GaeSla06,Wataru-etal17,Reynolds-etal17}, from intergalactic
medium \cite{Reynolds-etal17,Padovani21}, and when interacting
with strong magnetic fields in the vicinity of neutron stars \cite{HarLai06}%
. Theoretical aspects and experimental signatures of the synchrotron
radiation from astrophysical observations are discussed in many references; see, e.g., Refs. \cite{KelProAha15,BerPulVol09,StrOrlJaf11,KotAllLem11,Kenta18} and references therein. In laboratory
settings, synchrotron radiation plays a crucial role in the advancement of
various scientific fields, from exploring the atomic structure of advanced
materials for investigations in biological systems. The European Synchrotron
Radiation Facility (ESRF) \cite{ESRF} for instance produces ultra-brilliant
X-ray beams aiming to explore various phenomena under extreme conditions, as
recently detailed in Ref. \cite{Patrick-etal24}. The formula for the energy
rate emitted from an electron in a constant and homogeneous magnetic field
in the framework of classical electrodynamics was first presented in 1907 by
G. A. Schott \cite{Schotta,Schottb} and later expanded in his book in 1912 
\cite{Schott}. Years later, Schwinger \cite{Schwinger49} rederived the
classical energy rate emitted from the electron in a circular motion through
the \textquotedblleft source point-of-view\textquotedblright , which is
based on the rate at which the electron does work on the external field.
Comprehensive and extensive discussion on the theory of synchrotron
radiation can be found in a series of monographs; see, e.g., Refs. \cite%
{SokTe68,Bordovitsyn99,Wiedemann} and in the textbook \cite{Schwbook} as
well.

Besides synchrotron radiation, another fundamentally important class of
electromagnetic radiation arises when charged particles are accelerated by
external plane-wave fields. This type of radiation--which can be interpreted
as a scattering process between the charge (electron, for instance) and the
external plane-wave field--corresponds classically to the Thomson scattering 
\cite{Jacks99} or the Compton scattering \cite{ItzZub} in Quantum
Electrodynamics (QED). However, when the intensity of the external field is
sufficiently strong, the particle-field interaction enters a nonlinear
regime and gives rise to processes known as nonlinear Thomson scattering and
nonlinear Compton scattering. The theoretical foundations of QED with a
plane-wave background field was pioneered in the works by Reiss \cite%
{Reiss62}, Goldman \cite{Goldman64}, Brown \& Kibble \cite{BroKib64}, and
Nikishov \& Ritus \cite{NikRit64}. In Refs. \cite{BroKib64,NikRit64} the
authors employed the exact solutions of the Dirac equation in a plane-wave
field (Volkov solutions \cite{Volkov}) to calculate probabilities of
fundamental processes, such as one-photon emission by an electron and
probabilities of pair creation by a photon in such a background. Ritus \cite%
{Ritus85}, in particular, formalized these calculations within the Furry
representation of QED with external fields \cite{Furry51}, in which the
interaction with the plane-wave background is taken into account exactly.
These seminal works established the theoretical framework for fundamental
processes in QED in a plane-wave background, including the nonlinear Compton
scattering and the nonlinear Breit-Wheeler pair production--the creation of
electron-positron pair production by a photon in the plane-wave field.

While the effects predicted by these early theoretical works remained
experimentally inaccessible for decades, recent advancements in
ultra-intense laser facilities, such as the Europeans XFEL \cite{XFEL}, ELI 
\cite{ELI}, DESY \cite{DESY} and the Linac Coherent Laser Source \cite{LINAC}%
\ in the US, has sparked significant activity in both theoretical and
phenomenological studies of quantum processes in processes, including the
impact of the beam shape \cite{HeiSeiKam10,KraKam12} and radiation-reaction
effects in the spectrum of nonlinear Thomson scattering \cite%
{ShePia12,Piazza18,PiaAud21}, analysis of the locally-constant field
approximation (LCFA) \cite{PiaTamMeuKei18,IldKinSei19}, effects of the
electron wave packet \cite{AngMacPia16}, the role of photon polarization 
\cite{KinTan20}, and interference from multiple laser pulses \cite%
{IldKinTan20} in the nonlinear Compton scattering spectrum. Experimental
observation of the nonlinear Compton scattering was also reported in Ref. 
\cite{Bula-etal96}. Besides the nonlinear Compton scattering, various
aspects underlying the nonlinear Breit-Wheeler pair production in
plane-wave-like fields, such as realistic beam configurations, polarization
effects, pulse effects, and multi-pulse interactions, and
dynamical-assistance related effects were investigated in several
references, e.g. \cite%
{JanMul13,KraKa14,WuXue14,MeuHatKeiPia15,Otto-etal16,DiPiazza16,MeuKeiPia16,JanMul16,JanMul17,GolChaMul22,MahChaMul23}%
. Numerical analysis of the nonlinear Compton scattering and the
Breit-Wheeler process was also considered in some works, for instance, in
Refs. \cite{HarIldKin15,BlaSeiBulMar18,BlaSeiBulMar20}. An extensive
discussion of experimental, theoretical, and phenomenological aspects
related to quantum effects stemming from particle-field interaction in
strong external fields can be found in the review papers \cite%
{MouTajBul06,MarShu06,EhlKraKam09,Bulanov-etal11,PiaMulHatKei12,Blackburn20,GonBlaMarBul22,Fedotov-etal23,Sarri-etal25}
and pertinent references therein.

The classical expression for the energy rate emitted from an electron
interacting with a plane-wave field was presented in Landau and Lifshitz's
textbook \cite{LanLi71} and more systematically derived in a series of works
by Sokolov, Ternov, and collaborators in the late 1960s \cite%
{SokTerBagGalShu68,SokZhuKol69,SokGalKol71}. These results were subsequently
compiled in their book \cite{SokTe68}, and additional aspects are discussed
in Jackson's textbook \cite{Jacks99}. The classical energy rate discussed in
these references are calculated through the Umov-Poynting energy-flux vector 
\cite{Umov1874,Poynt1884} and based on the Heaviside-Poynting's theorem \cite%
{Poynt1884,Heavi1884}, which relies on a number of hypotheses as well
detailed in Stratton's textbook \cite{Strat41}. Another issue characteristic
of classical energy is the appearance of divergences associated with the
duration that sources interact with the external field. Depending on the
external field, the classical energy \cite{LanLi71,Jacks99}%
\begin{equation}
W_{\mathrm{cl}}=4\pi ^{2}\int \left\vert \mathbf{n}\times \left[ \mathbf{n}%
\times \mathbf{\tilde{j}}\left( k\right) \right] \right\vert ^{2}d\mathbf{k}%
\,,  \label{de2}
\end{equation}%
radiated from the current distribution $\mathbf{j}\left( x\right) $--whose
Fourier transform is \\ $\mathbf{\tilde{j}}\left( k\right) =\left( 2\pi \right)
^{-2}\int e^{ikx}\mathbf{j}\left( x\right) dx$--exhibit divergences when
calculating the integrals over $\mathbf{k}$. This issue was reported by
Nikishov and Ritus \cite{NikRit69} in the context of a point particle
interacting with a constant and uniform electric field. In this particular
example, Nikishov and Ritus concluded that the energy radiated from the
electron is divergent as it is perpetually accelerated by the field and
radiates energy at a constant rate. Nevertheless, the problem can be
circumvented if the external field switches on and off at remote
times or if the current distribution is exposed to the field over a finite
time interval, which amounts to splitting the integral over $t^{\prime }$ in
(\ref{de2}) into intervals where the particle interacts with the field and
where it is free. Yet, the latter contributions still require some sort of
regularization as pointed out in Jackson's textbook \cite{Jacks99}. The
quantum theory features similar problems, as discussed by Nikishov \cite%
{Nikishov71}\ in the context of photon emission from an electron in an
infinitely constant and homogeneous electric field\footnote{%
This process is analogous to nonlinear Compton scattering in a constant
electric field.}.

In this work, we employ a semiclassical method for calculating the
electromagnetic energy and the energy rate emitted from an electron in an
external plane-wave field. In this formulation, the current is treated
classically while the radiated electromagnetic field is quantum. The theory
is based on the evolution of the quantum state of the electromagnetic field
from an initial state without photons at time $t_{\mathrm{in}}$ to a state
with photons at time $t$. The corresponding transition probability has been
presented in detail in Refs. \cite{BagGiSF20,ShiLeBG21,TGJB}. Using such a
probability, we have shown in \cite{TGJB} that the total energy radiated
from the particle is analogous to the classical result (\ref{de2})%
\begin{equation}
W\left( \Delta t\right) =4\pi ^{2}\int \left\vert \mathbf{n}\times \left[ 
\mathbf{n}\times \mathbf{\tilde{j}}\left( k;\Delta t\right) \right]
\right\vert ^{2}d\mathbf{k}\,,  \label{de1}
\end{equation}%
with%
\begin{equation}
\mathbf{\tilde{j}}\left( k;\Delta t\right) =\frac{1}{\left( 2\pi \right) ^{2}%
}\int_{t_{\mathrm{in}}}^{t}dt^{\prime }e^{ik_{0}ct^{\prime }}\int e^{-i%
\mathbf{kr}}\mathbf{j}\left( t^{\prime },\mathbf{r}\right) d\mathbf{r}\,,
\label{de1b}
\end{equation}%
representing an \textit{incomplete} Fourier transform of the current
density. The finite integration range in this formula stems from the
probability of the process to occur within the transition interval $\Delta
t=t-t_{\mathrm{in}}$, which is inherently linked to the quantum description
of the radiation process \cite{BagGiSF20,ShiLeBG21,TGJB}. This feature
yields to a finite radiated energy, as discussed in Ref. \cite{TGJB} and in
the system under consideration below. The compatibility with the classical
radiation spectrum (\ref{de2}) is achieved in the limit where the quantum
transition interval approaches infinity, $\Delta t\rightarrow \infty $, as $%
\lim_{\Delta t\rightarrow \infty }\mathbf{\tilde{j}}\left( k;\Delta t\right)
=\mathbf{\tilde{j}}\left( k\right) $. Processes take place in the
four-dimensional Minkowski space-time with coordinates $x=\left( x^{\mu },\
\mu =0,i\right) =\left( ct,\mathbf{r}\right) $, $ct=x^{0}$, $\mathbf{r}%
=\left( x^{i},\ i=1,2,3\right) $, and metric tensor $\eta _{\mu \nu }=%
\mathrm{diag}\left( +1,-1,-1,-1\right) $.\ The Gauss system of units is used.

\section{Electromagnetic energies and rates radiated from a charged particle
in a plane wave field\label{Sec2}}

\subsection{General\label{Sec2.0}}

In the presence of an external monochromatic plane-wave field propagating
along the vector $\mathbf{n}_{\mathrm{w}}$, $\left\vert \mathbf{n}_{\mathrm{w%
}}\right\vert =1$, with wave four vector $k_{\mathrm{w}}^{\mu }=\omega _{%
\mathrm{w}}n_{\mathrm{w}}^{\mu }/c$, $n_{\mathrm{w}}^{\mu }=\left( 1,\mathbf{%
n}_{\mathrm{w}}\right) $, $k_{\mathrm{w}}^{0}=\left\vert \mathbf{k}_{\mathrm{%
w}}\right\vert =\omega _{\mathrm{w}}/c$, and angular frequency $\omega _{%
\mathrm{w}}$, the motion of a charged particle with charge $e$ (for an
electron $e=-e_{0}$, $e_{0}>0$) in the laboratory frame has the form \cite%
{BagGit2000}\footnote{%
Lorentz contraction between two arbitrary four vectors $A^{\mu }$ and $%
B^{\mu }$ are conveniently represented as $A_{\mu }B^{\mu }=\left( AB\right)
=A^{0}B^{0}-\mathbf{AB}$.}:%
\begin{eqnarray}
&&\mathbf{r}\left( \phi \right) =\underline{\mathbf{r}}+\frac{c}{p_{-}}\int
\left( \boldsymbol{\kappa }-\frac{e}{c}\mathbf{A}\right) d\phi +\mathbf{n}_{%
\mathrm{w}}\frac{c}{2p_{-}^{2}}\int \left[ m^{2}c^{2}+\left( \boldsymbol{%
\kappa }-\frac{e}{c}\mathbf{A}\right) ^{2}-p_{-}^{2}\right] d\phi \,,  \notag
\\
&&ct\left( \phi \right) =c\underline{t}+\frac{c}{2p_{-}^{2}}\int \left[
m^{2}c^{2}+\left( \boldsymbol{\kappa }-\frac{e}{c}\mathbf{A}\right)
^{2}+p_{-}^{2}\right] d\phi \,,\ \ \phi =\frac{\left( n_{\mathrm{w}}x\right) 
}{c}\,.  \label{3.1}
\end{eqnarray}%
Here, $A^{\mu }=A^{\mu }\left( \phi \right) =\left( A^{0}\left( \phi \right)
,\mathbf{A}\left( \phi \right) \right) $ is the electromagnetic potential of
the plane-wave, $\underline{x}^{\mu }=\left( c\underline{t},\underline{%
\mathbf{r}}\right) $ is the initial position of the particle in the
space-time, $\left( n_{\mathrm{w}}x\right) =n_{\mathrm{w}}^{\mu }x_{\mu
}=x^{0}-\mathbf{n}_{\mathrm{w}}\mathbf{r}$, and the vector $\boldsymbol{%
\kappa }=\left( \kappa _{x},\kappa _{y},\kappa _{z}\right) $ is an integral
of motion that is orthogonal to $\mathbf{n}_{\mathrm{w}}$, $\boldsymbol{%
\kappa }\mathbf{n}_{\mathrm{w}}=0$. Additionally, $p_{-}=\left( n_{\mathrm{w}%
}P\right) =P^{0}-\left( \mathbf{n}_{\mathrm{w}}\mathbf{P}\right) $ is
another integral of motion, where $P^{\mu }=\left( P^{0},\mathbf{P}\right) $
is the kinetic four-momentum of the particle,%
\begin{eqnarray}
&&P^{0}=\frac{\mathcal{E}-A^{0}}{c}=\gamma mc=\frac{m^{2}c^{2}+\left( 
\boldsymbol{\kappa }-\frac{e}{c}\mathbf{A}\right) ^{2}+p_{-}^{2}}{2p_{-}}\,,
\notag \\
&&\mathbf{P}=\gamma mc\boldsymbol{\beta }=\boldsymbol{\kappa }-\frac{e}{c}%
\mathbf{A}+\mathbf{n}_{\mathrm{w}}\frac{m^{2}c^{2}+\left( \boldsymbol{\kappa 
}-\frac{e}{c}\mathbf{A}\right) ^{2}-p_{-}^{2}}{2p_{-}}\,,  \label{P}
\end{eqnarray}%
$\mathcal{E}$ and $\gamma $ are, respectively, its energy and Lorentz
factor, $\gamma =\left( 1-\boldsymbol{\beta }^{2}\right) ^{-1/2}$, $%
\boldsymbol{\beta }=c^{-1}d\mathbf{r}\left( t\right) /dt$. It is also useful
to express the momentum of the particle in a covariant form%
\begin{equation}
P^{\mu }=P^{\mu }\left( \phi \right) =q^{\mu }-\frac{e}{c}A^{\mu }+\frac{%
n^{\mu }}{2p_{-}}\left[ \frac{2e}{c}\left( qA\right) -\frac{e^{2}}{c^{2}}%
A^{2}\right] \,,  \label{P2}
\end{equation}%
where $q^{\mu }=\left( q^{0},\mathbf{q}\right) $, is the so-called
quasi-momentum%
\begin{equation}
q^{0}=\frac{m^{2}c^{2}+\boldsymbol{\kappa }^{2}+p_{-}^{2}}{2p_{-}}\,,\ \ 
\mathbf{q}=\boldsymbol{\kappa }+\mathbf{n}_{\mathrm{w}}\frac{m^{2}c^{2}+%
\boldsymbol{\kappa }^{2}-p_{-}^{2}}{2p_{-}}\,,  \label{P3}
\end{equation}%
as it satisfies the customary energy-momentum relation $q_{0}^{2}-\mathbf{q}%
^{2}=P_{0}^{2}-\mathbf{P}^{2}=m^{2}c^{2}$.

To effectively calculate the electromagnetic energy (\ref{de1}) and rate
emitted from the particle, it is helpful to convert the time integral in (%
\ref{de1b}) into an integral over the phase $\phi $ as electromagnetic
potentials of plane-wave fields depend exclusively on the phase $\phi $.
This can be achieved via the substitution $\varphi =\omega _{\mathrm{w}}\phi
=\omega _{\mathrm{w}}\left( t-\mathbf{n}_{\mathrm{w}}\mathbf{r}\left(
t\right) /c\right) $, such that integrals over time are transformed into
integrals over the phase as follows%
\begin{equation*}
\int_{t_{\mathrm{in}}}^{t}dt^{\prime }\rightarrow \frac{1}{\omega _{\mathrm{w%
}}}\int_{\varphi _{\mathrm{in}}}^{\varphi }\frac{d\varphi ^{\prime }}{1-%
\mathbf{n}_{\mathrm{w}}\boldsymbol{\beta }\left( t^{\prime }\right) }\,,\ \
\varphi _{\mathrm{in}}=\varphi \left( t_{\mathrm{in}}\right) ,\ \ \varphi
=\varphi \left( t\right) \,.
\end{equation*}%
Furthermore, the calculation of the energy radiated by the charge in the
given background can be simplified by using the conservation of the electric
charge $\partial _{\mu }j^{\mu }\left( x\right) =0$ \cite{Jacks99,LanLi71}.
This allows us to substitute the double cross product (\ref{de1}) with a
Minkowski product of four currents,%
\begin{equation}
W\left( \Delta \varphi \right) =-4\pi ^{2}\int \left\vert \tilde{j}^{\mu
}\left( k\mathbf{;}\Delta \varphi \right) \right\vert ^{2}d\mathbf{k}\,,
\label{de3}
\end{equation}%
where%
\begin{equation}
\tilde{j}^{\mu }\left( k;\Delta \varphi \right) =\frac{e}{4\pi ^{2}}%
\int_{\varphi _{\mathrm{in}}}^{\varphi }\frac{P^{\mu }\left( \varphi
^{\prime }\right) }{\left( k_{\mathrm{w}}P\left( \varphi ^{\prime }\right)
\right) }\exp \left[ i\int_{\varphi _{0}^{\prime }}^{\varphi ^{\prime }}%
\frac{\left( kP\left( \varphi ^{\prime \prime }\right) \right) }{\left( k_{%
\mathrm{w}}P\left( \varphi ^{\prime \prime }\right) \right) }d\varphi
^{\prime \prime }\right] d\varphi ^{\prime }\,,  \label{de1c}
\end{equation}%
is the Fourier transform of point particle $j^{\mu }\left( x\right) =\left(
c\rho \left( x\right) ,c\rho \left( x\right) \boldsymbol{\beta }\left(
t\right) \right) $, $\rho \left( x\right) =e\delta ^{3}\left( \mathbf{r}-%
\mathbf{r}\left( t\right) \right) $, with momentum (\ref{P2}). Here, $k^{\mu
}=\omega n^{\mu }/c$,$\ n^{\mu }=\left( 1,\mathbf{n}\right) $, $%
k^{0}=\left\vert \mathbf{k}\right\vert =\omega /c$, denotes the wave four
vector of the radiated field, $\Delta \varphi =\varphi -\varphi _{\mathrm{in}%
}$, and $\varphi _{0}^{\prime }$ is an arbitrary constant. In the limit the
quantum transition interval approaches infinity, $\Delta t\rightarrow \infty 
$, which implies in $\Delta \varphi \rightarrow \infty $, we recover the
classical expression for the current distribution%
\begin{equation}
\tilde{j}^{\mu }\left( k\mathbf{;}\infty \right) =\tilde{j}^{\mu }\left(
k\right) =\frac{e}{4\pi ^{2}}\int_{-\infty }^{\infty }\frac{P^{\mu }\left(
\varphi ^{\prime }\right) }{\left( k_{\mathrm{w}}P\right) }\exp \left[
i\int_{\varphi _{0}^{\prime }}^{\varphi ^{\prime }}\frac{\left( kP\left(
\varphi ^{\prime \prime }\right) \right) }{\left( k_{\mathrm{w}}P\right) }%
d\varphi ^{\prime \prime }\right] d\varphi ^{\prime }\,,  \label{de1d}
\end{equation}%
and for the classical energy $W_{\mathrm{cl}}$, when (\ref{de1d}) is
substituted in (\ref{de3}). We call this limit the classical limit for
convenience. The calculation of the semiclassical energy (\ref{de3}) and
energy rate emitted from an electron in a circularly polarized and linearly
polarized plane wave fields are discussed in the sections below.

\subsection{Circularly-polarized external field\label{Sec2.1}}

In this section, we consider a circularly polarized plane wave field
propagating along the positive $z$-axis of the laboratory frame, $\mathbf{n}%
_{\mathrm{w}}=\left( 0,0,1\right) $. The vector potential of the field can
be written in the form,%
\begin{equation}
A^{\mu }\left( \varphi \right) =-\frac{cE_{0}}{\omega _{\mathrm{w}}}\left(
0,\sin \varphi ,-\varkappa \cos \varphi ,0\right) \,,\ \ \varphi =\omega _{%
\mathrm{w}}\phi \,,  \label{A}
\end{equation}%
where $\varkappa =+1\left( -1\right) $ describes a right(left)-handed
polarized plane wave, and $E_{0}$ denotes the amplitude of the electric
field. Plugging the potential (\ref{A}) into Eqs. (\ref{3.1}) and (\ref{P})
we obtain%
\begin{eqnarray}
x\left( \varphi \right) &=&\underline{x}+\frac{c\kappa _{x}\varphi }{\omega
_{\mathrm{w}}p_{-}}-\frac{ceE_{0}}{\omega _{\mathrm{w}}^{2}p_{-}}\cos
\varphi \,,  \notag \\
y\left( \varphi \right) &=&\underline{y}+\frac{c\kappa _{y}\varphi }{\omega
_{\mathrm{w}}p_{-}}-\frac{\varkappa ceE_{0}}{\omega _{\mathrm{w}}^{2}p_{-}}%
\sin \varphi \,,  \notag \\
z\left( \varphi \right) &=&\underline{z}+\left[ 2\frac{q_{z}-\kappa _{z}}{%
p_{-}}+\left( \frac{eE_{0}}{p_{-}\omega _{\mathrm{w}}}\right) ^{2}\right] 
\frac{\varphi }{2k_{\mathrm{w}}^{0}} -\frac{ceE_{0}}{\omega _{\mathrm{w}}^{2}p_{-}^{2}}\left( \kappa _{x}\cos
\varphi +\varkappa \kappa _{y}\sin \varphi \right) \,,  \notag \\
ct\left( \varphi \right) &=&c\underline{t}+\left[ \frac{2q^{0}}{p_{-}}%
+\left( \frac{eE_{0}}{p_{-}\omega _{\mathrm{w}}}\right) ^{2}\right] \frac{%
\varphi }{2k_{\mathrm{w}}^{0}} -\frac{ceE_{0}}{\omega _{\mathrm{w}}^{2}p_{-}^{2}}\left( \kappa _{x}\cos
\varphi +\varkappa \kappa _{y}\sin \varphi \right) \,,  \label{phsol}
\end{eqnarray}%
and%
\begin{eqnarray}
P_{x} &=&\kappa _{x}+\frac{eE_{0}}{\omega _{\mathrm{w}}}\sin \varphi \,,\ \
P_{y}\left( \varphi \right) =\kappa _{y}-\varkappa \frac{eE_{0}}{\omega _{%
\mathrm{w}}}\cos \varphi \,,  \notag \\
P_{z} &=&q_{z}+\frac{1}{2p_{-}}\left( \frac{eE_{0}}{\omega _{\mathrm{w}}}%
\right) ^{2}+\frac{eE_{0}}{p_{-}\omega _{\mathrm{w}}}\left( \kappa _{x}\sin
\varphi -\varkappa \kappa _{y}\cos \varphi \right) \,,  \notag \\
P^{0} &=&q^{0}+\frac{1}{2p_{-}}\left( \frac{eE_{0}}{\omega _{\mathrm{w}}}%
\right) ^{2}+\frac{eE_{0}}{p_{-}\omega _{\mathrm{w}}}\left( \kappa _{x}\sin
\varphi -\varkappa \kappa _{y}\cos \varphi \right) \,.  \label{phsol-b}
\end{eqnarray}%
We stress that the time when the initial conditions for the particle's
motion lies within the interval during which the radiation is
produced, $\left[ t_{\mathrm{in}},t\right] $, in accordance with Eq. (\ref%
{de1})\textrm{.}

To simplify subsequent calculations, we can set $\boldsymbol{\kappa }=%
\mathbf{0}$ without loss of generality, and assume that the particle is at
the origin when $\tau =\tau _{0}$ (or, equivalently, when $\varphi =0$). In
this case, the trajectory of the particle and its momentum read%
\begin{eqnarray}
&&x\left( \varphi \right) =r_{\perp }\cos \varphi \,,\ \ P_{x}\left( \varphi
\right) =-mc\xi \sin \varphi \,,  \notag \\
&&y\left( \varphi \right) =\varkappa r_{\perp }\sin \varphi \,,\ \
P_{y}\left( \varphi \right) =\varkappa mc\xi \cos \varphi \,,  \notag \\
&&z\left( \varphi \right) =\frac{P_{z}}{k_{\mathrm{w}}^{0}p_{-}}\varphi \,,\
\ P_{z}=q_{z}+\frac{p_{-}}{2}\left( \frac{mc}{p_{-}}\xi \right) ^{2}\,, 
\notag \\
&&ct\left( \varphi \right) =\frac{P_{0}}{k_{\mathrm{w}}^{0}p_{-}}\varphi
\,,\ \ P_{0}=q_{0}+\frac{p_{-}}{2}\left( \frac{mc}{p_{-}}\xi \right) ^{2}\,,
\label{phsol2}
\end{eqnarray}%
where $r_{\perp }$ is the radius of the particle's orbit in the plane
perpendicular to $\mathbf{n}_{\mathrm{w}}$ and $\xi $ is the so-called
classical nonlinearity parameter\footnote{%
In the literature, \textquotedblleft $a_{0}$\textquotedblright\ is also used
to represent this parameter. See e.g. \cite{GonBlaMarBul22}.}:%
\begin{equation}
r_{\perp }=\frac{e_{0}}{c}\frac{\sqrt{-\left( F_{\mu \nu }P^{\nu }\right)
^{2}}}{\left( k_{\mathrm{w}}P\right) ^{2}}=\frac{ce_{0}E_{0}}{\omega _{%
\mathrm{w}}^{2}p_{-}}\,,\ \ \xi =\frac{e_{0}}{mc^{2}}\frac{\sqrt{-\left(
F_{\mu \nu }P^{\nu }\right) ^{2}}}{\left( k_{\mathrm{w}}P\right) }=\frac{%
e_{0}E_{0}}{mc\omega _{\mathrm{w}}}\,.  \label{radius}
\end{equation}%
The classical nonlinearity parameter, as extensively discussed in the
literature, e.g. in Refs. \cite%
{NikRit64,NikRit67,HeiIld09,RufVerXue10,PiaMulHatKei12,GonBlaMarBul22,Fedotov-etal23}%
, quantifies the coupling between charge and the external field. It can be
interpreted in several ways: as the work exerted by the external field on a
charge over the electron Compton wavelength $\lambdabar_{\mathrm{C}}=\hslash/mc$, in units of the external photon
energy $\hslash \omega _{\mathrm{w}}$; as the ratio between the transversal
energy of the electron in the field with its rest energy, or as the
amplitude of the wave four-potential in units of $mc\omega _{\mathrm{w}%
}/e_{0}$. Explicitly:%
\begin{equation}
\xi =\frac{e_{0}E_{0}\lambdabar_{\mathrm{C}}}{\hslash \omega _{\mathrm{w}}}=%
\frac{c\left\vert \mathbf{P}_{\perp }\right\vert }{mc^{2}}=\frac{e_{0}\sqrt{%
-A^{2}}}{mc^{2}}\,,\ \
\left\vert \mathbf{P}_{\perp }\right\vert =\sqrt{P_{x}^{2}+P_{y}^{2}}\,,
\label{a0}
\end{equation}%
In particular, if the field amplitude is equal to the Schwinger critical value, 
$E_{\mathrm{cr}}=m^{2}c^{3}/e_{0}\hslash $, the classical nonlinearity
parameter becomes the ratio between the electron rest energy and the
(external) photon energy,%
\begin{equation}
\xi _{\mathrm{cr}}=\left. \xi \right\vert _{E_{0}=E_{\mathrm{cr}}}=\frac{%
mc^{2}}{\hslash \omega _{\mathrm{w}}}\,.  \label{a0cr}
\end{equation}%
In addition to $\xi $, photon emission characteristics are also expressed
through a parameter known as the quantum nonlinearity parameter $\chi _{%
\mathrm{e}}$,%
\begin{equation}
\chi _{\mathrm{e}}=\frac{\sqrt{-\left( F^{\mu \nu }P_{\nu }\right) ^{2}}}{%
mcE_{\mathrm{cr}}}=\frac{E_{0}}{E_{\mathrm{cr}}}\frac{p_{-}}{mc}\,,
\label{chie}
\end{equation}%
which quantifies the significance of quantum corrections to the radiation
emitted from the charge's motion. A detailed discussion of the values of
these parameters across various experiments can be found in several references; see e.g., Refs. \cite{HeiIld09,PiaMulHatKei12,GonBlaMarBul22,Fedotov-etal23,Blackburn20,Sarri-etal25}
and references therein.

The solutions presented above allow us to calculate electromagnetic energies
and rates from an electron in a circularly-polarized plane wave field. By
plugging the momenta (\ref{phsol2}) into (\ref{de1c}) and performing a
subsequent change of variable,%
\begin{equation}
\Phi ^{\prime }=\varphi ^{\prime }-\varkappa \varphi _{\gamma }+\pi /2\,,
\label{var}
\end{equation}%
we can present the Fourier transform (\ref{de1c}) of the current density of an electron moving in the field in the form,%
\begin{eqnarray}
\tilde{j}^{0}\left( k;\Delta \varphi \right) &=&-\frac{e^{iC^{\prime }}}{%
4\pi ^{2}}\frac{e_{0}}{k_{\mathrm{w}}^{0}p_{-}}\left[ q_{0}+\frac{p_{-}}{2}%
\left( \frac{mc}{p_{-}}\xi \right) ^{2}\right] \int_{\Phi _{\mathrm{in}%
}}^{\Phi }e^{i\left( \eta \Phi ^{\prime }-\mu \sin \Phi ^{\prime }\right)
}d\Phi ^{\prime }\,,  \notag \\
\tilde{j}_{x}\left( k;\Delta \varphi \right) &=&\frac{e^{iC^{\prime }}}{4\pi
^{2}}e_{0}r_{\perp }\int_{\Phi _{\mathrm{in}}}^{\Phi }\left( \varkappa \sin
\varphi _{\gamma }\sin \Phi ^{\prime }-\cos \varphi _{\gamma }\cos \Phi
^{\prime }\right) e^{i\left( \eta \Phi ^{\prime }-\mu \sin \Phi ^{\prime
}\right) }d\Phi ^{\prime }\,,  \notag \\
\tilde{j}_{y}\left( k;\Delta \varphi \right) &=&\frac{e^{iC^{\prime }}}{4\pi
^{2}}e_{0}r_{\perp }\int_{\Phi _{\mathrm{in}}}^{\Phi }\left( -\varkappa \cos
\varphi _{\gamma }\sin \Phi ^{\prime }-\sin \varphi _{\gamma }\cos \Phi
^{\prime }\right) e^{i\left( \eta \Phi ^{\prime }-\mu \sin \Phi ^{\prime
}\right) }d\Phi ^{\prime }\,,  \notag \\
\tilde{j}_{z}\left( k;\Delta \varphi \right) &=&-\frac{e^{iC^{\prime }}}{%
4\pi ^{2}}\frac{e_{0}}{k_{\mathrm{w}}^{0}p_{-}}\left[ q_{z}+\frac{p_{-}}{2}%
\left( \frac{mc}{p_{-}}\xi \right) ^{2}\right] \int_{\Phi _{\mathrm{in}%
}}^{\Phi }e^{i\left( \eta \Phi ^{\prime }-\mu \sin \Phi ^{\prime }\right)
}d\Phi ^{\prime }\,,  \label{cur1}
\end{eqnarray}%
where $C^{\prime }$ is an unimportant constant phase, $\Phi =\Phi \left(
\varphi \right) $, $\Phi _{\mathrm{in}}=\Phi \left( \varphi _{\mathrm{in}%
}\right) $, and%
\begin{equation}
\eta =\frac{k^{0}}{k_{\mathrm{w}}^{0}}\frac{P^{0}-P_{z}\cos \theta _{\gamma }%
}{p_{-}}\,,\ \ \mu =\frac{k^{0}}{k_{\mathrm{w}}^{0}}\frac{mc}{p_{-}}\xi \sin
\theta _{\gamma }\,.  \label{v1b}
\end{equation}%
Next, expanding the exponentials in (\ref{cur1}) in terms of Bessel
functions with the aid of the identities \cite{Watson,Watsonb},%
\begin{eqnarray}
&&e^{-i\alpha \sin \tau }=\sum_{n=-\infty }^{+\infty }J_{n}\left( \alpha
\right) e^{-in\tau }\,,\ \ e^{-i\alpha \sin \tau }\cos \tau =\sum_{n=-\infty
}^{+\infty }\frac{n}{\alpha }J_{n}\left( \alpha \right) e^{-in\tau }\,, 
\notag \\
&&e^{-i\alpha \sin \tau }\sin \tau =i\sum_{n=-\infty }^{+\infty
}J_{n}^{\prime }\left( \alpha \right) e^{-in\tau }\ ,\ \ J_{n}^{\prime
}\left( \alpha \right) =\frac{dJ_{n}\left( \alpha \right) }{d\alpha }\,,
\label{bessel}
\end{eqnarray}%
we can present the currents (\ref{cur1}) in the form%
\begin{eqnarray}
\tilde{j}^{0}\left( k;\Delta \varphi \right) &=&-\frac{e^{iC^{\prime }}e_{0}%
}{2\pi ^{2}}\left[ \frac{q_{0}}{k_{\mathrm{w}}^{0}p_{-}}+\frac{1}{2k_{%
\mathrm{w}}^{0}}\left( \frac{mc}{p_{-}}\xi \right) ^{2}\right]
\sum_{n=-\infty }^{+\infty }J_{n}\left( \mu \right) \int_{\Phi _{\mathrm{in}%
}}^{\Phi }e^{i\left( \eta -n\right) \Phi ^{\prime }}d\Phi ^{\prime }\,, 
\notag \\
\tilde{j}_{x}\left( k;\Delta \varphi \right) &=&\frac{e^{iC^{\prime
}}e_{0}r_{\perp }}{2\pi ^{2}}\sum_{n=-\infty }^{+\infty }\left( -\frac{n}{%
\mu }J_{n}\left( \mu \right) \cos \varphi _{\gamma }+i\varkappa
J_{n}^{\prime }\left( \mu \right) \sin \varphi _{\gamma }\right) \int_{\Phi
_{\mathrm{in}}}^{\Phi }e^{i\left( \eta -n\right) \Phi ^{\prime }}d\Phi
^{\prime }\,,  \notag \\
\tilde{j}_{y}\left( k;\Delta \varphi \right) &=&\frac{e^{iC^{\prime
}}e_{0}r_{\perp }}{2\pi ^{2}}\sum_{n=-\infty }^{+\infty }\left( -i\varkappa
J_{n}^{\prime }\left( \mu \right) \cos \varphi _{\gamma }-\frac{n}{\mu }%
J_{n}\left( \mu \right) \sin \varphi _{\gamma }\right) \int_{\Phi _{\mathrm{%
in}}}^{\Phi }e^{i\left( \eta -n\right) \Phi ^{\prime }}d\Phi ^{\prime }\,, 
\notag \\
\tilde{j}_{z}\left( k;\Delta \varphi \right) &=&-\frac{e^{iC^{\prime }}e_{0}%
}{2\pi ^{2}}\left[ \frac{q_{z}}{k_{\mathrm{w}}^{0}p_{-}}+\frac{1}{2k_{%
\mathrm{w}}^{0}}\left( \frac{mc}{p_{-}}\xi \right) ^{2}\right]
\sum_{n=-\infty }^{+\infty }J_{n}\left( \mu \right) \int_{\Phi _{\mathrm{in}%
}}^{\Phi }e^{i\left( \eta -n\right) \Phi ^{\prime }}d\Phi ^{\prime }\,.
\label{ch7}
\end{eqnarray}

Squaring the modulus of these currents and restoring the original variable
through Eq. (\ref{var}), we transform the remaining integrals over $\mathbf{k%
}$ in (\ref{de3}) in spherical coordinates, $d\mathbf{k}=c^{-3}\omega
^{2}d\omega d\Omega $, $d\Omega =\sin \theta _{\gamma }d\theta _{\gamma
}d\varphi _{\gamma }$, and integrate over the polar angle $\varphi _{\gamma
} $ to realize that the energy radiated from one photon (\ref{de3}) admits
the form,%
\begin{eqnarray}
W\left( \Delta \varphi \right) &=&\frac{e_{0}^{2}}{\pi c^{3}}\sum_{n=-\infty
}^{+\infty }\int_{0}^{\infty }d\omega \omega ^{2}\int_{0}^{\pi }\sin \theta
_{\gamma }T_{n}\left( \eta ,\Delta \varphi \right)  \notag \\
&\times &\left[ \left( \frac{n^{2}}{\mu ^{2}}-\frac{1}{\xi ^{2}}-1\right)
r_{\perp }^{2}\left\vert J_{n}\left( \mu \right) \right\vert ^{2}+r_{\perp
}^{2}\left\vert J_{n}^{\prime }\left( \mu \right) \right\vert ^{2}\right]
d\theta _{\gamma }\,,  \label{we}
\end{eqnarray}%
where the function $T_{n}\left( \eta ,\Delta \varphi \right) $%
\begin{equation}
T_{n}\left( \eta ,\Delta \varphi \right) =\frac{1-\cos \left[ \left( \eta
-n\right) \Delta \varphi \right] }{\left( \eta -n\right) ^{2}}\,,\ \ \Delta
\varphi =\varphi -\varphi _{\mathrm{in}}\,,  \label{we2}
\end{equation}%
encloses the time through which the radiation is formed, $\Delta t=t-t_{%
\mathrm{in}}$. Equation (\ref{we}) describes the total energy radiated from
an electron interacting with a circularly polarized plane wave field.
Although it has been derived within the semiclassical approach, this result
is classical as it does not feature quantities inherent to quantum theory,
such as the fine structure constant $\alpha$ or the quantum nonlinearity parameter (%
\ref{chie}). However, in contrast to the classical energy, our result (\ref%
{we}) is finite because the radiation is generated within the finite
transition interval $\Delta t$. To illustrate this, we first note that the
function (\ref{we2}) is undefined in the limit\footnote{%
or, equivalently, in the limit $\Delta t=t-t_{\mathrm{in}}\rightarrow \infty 
$.} $\Delta \varphi \rightarrow \infty $ but it is sharply peaked at the
resonance frequency $\omega _{\mathrm{res}}$%
\begin{equation}
\omega _{\mathrm{res}}\equiv n\omega _{\mathrm{r}}\,,\ \ \omega _{\mathrm{r}%
}=\omega _{\mathrm{w}}\frac{p_{-}}{P_{0}-P_{z}\cos \theta _{\gamma }}\,,
\label{we3}
\end{equation}%
with a characteristic width $\Delta \omega $ inversely proportional to the
quantum transition time:%
\begin{equation}
\Delta \omega =\frac{\omega _{\mathrm{r}}}{\omega _{\mathrm{w}}}\frac{P_{0}}{%
p_{-}}\frac{4\pi }{\Delta t}\,.  \label{we4}
\end{equation}%
Consequently, in this limit, the most significant contribution to the energy (\ref{we})
comes from a narrow bandwidth centered at the resonance frequency $%
\omega _{\mathrm{res}}$,%
\begin{eqnarray}
W\left( \Delta \varphi \right) &\approx &\frac{e_{0}^{2}}{\pi c^{3}}%
\sum_{n=0}^{\infty }\int_{0}^{\pi }d\theta _{\gamma }\sin \theta _{\gamma
}\int_{\omega _{\mathrm{res}}-\Delta \omega /2}^{\omega _{\mathrm{res}%
}+\Delta \omega /2}\omega ^{2}T_{n}\left( \eta ,\Delta \varphi \right) 
\notag \\
&\times &\left[ \left( \frac{n^{2}}{\mu ^{2}}-\frac{1}{\xi ^{2}}-1\right)
r_{\perp }^{2}\left\vert J_{n}\left( \mu \right) \right\vert ^{2}+r_{\perp
}^{2}\left\vert J_{n}^{\prime }\left( \mu \right) \right\vert ^{2}\right]
d\omega \,.  \label{we5}
\end{eqnarray}%
We note that the sum over negative integers vanishes identically since $%
\omega _{\mathrm{r}}$ is nonnegative and the integration over $\omega $ is
performed through a positive interval. While the latter integral cannot be
calculated exactly, it is possible to estimate an upper bound for the energy
if we confine ourselves to the limit of large $\Delta t$: expanding the
function (\ref{we2}) around the resonance frequency%
\begin{equation}
T_{n}\left( \eta ,\Delta \varphi \right) =\frac{\left( \Delta \varphi
\right) ^{2}}{2}+O\left( \Delta t^{-4}\right) \,,  \label{we5b}
\end{equation}%
neglecting terms of order $\Delta t^{-4}$ and evaluating the integrand of (%
\ref{we5}) at the resonance frequency, we obtain an upper bound for the
energy spectrum $W\left( \Delta \varphi \right) _{\max }$:%
\begin{eqnarray}
W\left( \Delta \varphi \right) _{\max } &\approx &2\Delta t\frac{\omega _{%
\mathrm{w}}^{2}e_{0}^{2}}{c}\left( \frac{p_{-}}{P_{0}}\right)
^{2}\sum_{n=1}^{\infty }n^{2}\int_{0}^{\pi }d\theta _{\gamma }\frac{\sin
\theta _{\gamma }}{\left( 1-\beta _{\parallel }\cos \theta _{\gamma }\right)
^{3}}  \notag \\
&\times &\left[ \left( \frac{\beta _{\parallel }-\cos \theta _{\gamma }}{%
\sin \theta _{\gamma }}\right) ^{2}J_{n}^{2}\left( \frac{n\left\vert 
\boldsymbol{\beta }_{\perp }\right\vert \sin \theta _{\gamma }}{1-\beta
_{\parallel }\cos \theta _{\gamma }}\right) +\boldsymbol{\beta }_{\perp
}^{2}J_{n}^{\prime 2}\left( \frac{n\left\vert \boldsymbol{\beta }_{\perp
}\right\vert \sin \theta _{\gamma }}{1-\beta _{\parallel }\cos \theta
_{\gamma }}\right) \right] \,,  \label{we6}
\end{eqnarray}%
Thus, the energy emitted from the electron (\ref{we}) is less than the
estimate above,\\ $W\left( \Delta \varphi \right) <W\left( \Delta \varphi
\right) _{\max }$. In the derivation of Eq. (\ref{we6}) we performed the substitutions%
\begin{equation}
\beta _{\parallel }=\frac{P_{z}}{P_{0}}\,,\ \ \left\vert \boldsymbol{\beta }%
_{\perp }\right\vert =\frac{e_{0}E_{0}}{\omega _{\mathrm{w}}P_{0}}\,,
\label{we6b}
\end{equation}%
that follow from the equations of motion (\ref{P}), (\ref{phsol2}).

The maximum energy (\ref{we6}) has two key characteristics. First, it is
linearly proportional to the time interval during which radiation is
produced, $\Delta t$. This indicates that the electromagnetic energy emitted
by the particle diverges if it perpetually interacts with the external field%
\footnote{%
Recall that the interaction time lies within the interval $\Delta t$, in
accordance with Eq. (\ref{de1}).}. This type of divergence has been reported
in the literature in other instances, such as when a charged particle is
linearly accelerated by a constant electric field \cite{NikRit69,TGJB}. To
our knowledge, this divergence has not been explicitly discussed in the
literature for the case under consideration\footnote{%
In Ref. \cite{Ritus85}, Ritus discussed the divergence of the energy in a
linearly polarized plane wave field. In the following section, we comment on
the absence of divergence in this case within the semiclassical formulation.}%
. We believe that the lack of discussion on this topic stems from the fact
that the classical electromagnetic energy radiated from a point particle--as
derived from Heaviside-Poynting's theorem \cite{Heavi1884,Poynt1884}--is
based on the integration of the energy rate over an infinite time interval.
When converted into an integral over the frequency of the radiation field
through Parseval's theorem, the resulting energy spectrum typically exhibits
divergences if the external field extends indefinitely in space, as it
continuously accelerates the particle. Consequently, there are difficulties
in defining the classical electromagnetic energy in these cases.
Nevertheless, it is possible to consider a situation where the particle
interacts with the external field during a finite interval within the
classical theory, as pointed out in Jackson's textbook \cite{Jacks99} and
studied for example in Ref. \cite{Sprangle93}. While this procedure
regularizes the energy, the authors of \cite{Sprangle93} did not discuss the
absence of divergence of the energy in relation to the finite time
duration. As discussed in our previous work \cite{TGJB}, the semiclassical
energy is finite because the radiation is generated during the quantum
transition interval $\Delta t$. Since the transition between quantum states
occurs within this interval, it naturally introduces a regularization to the
problem and eliminates the aforementioned divergence.

The second key characteristic encoded in (\ref{we6}) is that it enables us
to estimate an upper bound for the energy rate,%
\begin{equation}
w_{\max }=\frac{W\left( \Delta \varphi \right) _{\max }}{\Delta t}=2w_{%
\mathrm{cl}}\,,  \label{we7}
\end{equation}%
which, except by a factor of $2$, coincides with the classical rate at which
the energy is emitted from the electron in a circularly polarized plane wave
field:%
\begin{eqnarray}
w_{\mathrm{cl}} &=&\frac{\omega _{\mathrm{w}}^{2}e_{0}^{2}}{c}\left( \frac{%
p_{-}}{P_{0}}\right) ^{2}\int_{0}^{\pi }d\theta _{\gamma }\frac{\sin \theta
_{\gamma }}{\left( 1-\beta _{\parallel }\cos \theta _{\gamma }\right) ^{3}} 
\notag \\
&\times &\sum_{n=1}^{+\infty }n^{2}\left[ \left( \frac{\cos \theta _{\gamma
}-\beta _{\parallel }}{\sin \theta _{\gamma }}\right) ^{2}J_{n}^{2}\left( 
\frac{n\left\vert \boldsymbol{\beta }_{\perp }\right\vert \sin \theta
_{\gamma }}{1-\beta _{\parallel }\cos \theta _{\gamma }}\right) +\boldsymbol{%
\beta }_{\perp }^{2}J_{n}^{\prime 2}\left( \frac{n\left\vert \boldsymbol{%
\beta }_{\perp }\right\vert \sin \theta _{\gamma }}{1-\beta _{\parallel
}\cos \theta _{\gamma }}\right) \right] \,.  \label{pc1}
\end{eqnarray}%
This expression is the analog of Schott's formula \cite{Schott} for an
electron moving in a circularly polarized plane wave field; see e.g., Ref. 
\cite{SokTe68} and Eq. (\ref{schott}) below. This result can also be derived
from the semiclassical energy rate. Differentiating the function (\ref{we2})
with respect to time,%
\begin{equation*}
\frac{\partial }{\partial t}T_{n}\left( \eta ,\Delta \varphi \right) =\frac{%
p_{-}\omega _{\mathrm{w}}}{P_{0}}\frac{\sin \left[ \left( \eta -n\right)
\Delta \varphi \right] }{\left( \eta -n\right) }\,,
\end{equation*}%
we discover that the energy rate has the following form:%
\begin{eqnarray}
w\left( \Delta \varphi \right) &=&\frac{\partial }{\partial t}W\left( \Delta
\varphi \right) =\frac{p_{-}}{P_{0}}\frac{\omega _{\mathrm{w}}e_{0}^{2}}{\pi
c^{3}}\sum_{n=-\infty }^{+\infty }\int_{0}^{\infty }d\omega \omega
^{2}\int_{0}^{\pi }d\theta _{\gamma }\sin \theta _{\gamma }  \notag \\
&\times &\left[ \left( \frac{n^{2}}{\mu ^{2}}-\frac{1}{\xi ^{2}}-1\right)
r_{\perp }^{2}\left\vert J_{n}\left( \mu \right) \right\vert ^{2}+r_{\perp
}^{2}\left\vert J_{n}^{\prime }\left( \mu \right) \right\vert ^{2}\right] 
\frac{\sin \left[ \left( \eta -n\right) \Delta \varphi \right] }{\left( \eta
-n\right) }\,.  \label{pp1}
\end{eqnarray}%
This formula corresponds to the rate at which one photon is emitted from the
electron within the quantum radiation interval $\Delta t$. By using the
identities (\ref{we6b}) and applying the well-known limit%
\begin{equation}
\lim_{\Delta \varphi \rightarrow \infty }\frac{2\sin \left( s\Delta \varphi
\right) }{s}=2\pi \delta \left( s\right) \,,  \label{pp0}
\end{equation}%
we recover the classical energy rate (\ref{pc1}) from Eq. (\ref{pp1}), i.e., 
$w\left( \infty \right) =w_{\mathrm{cl}}$.

To conclude this section, it is important to discuss the energy radiated
from the electron in a frame where it is, on average, at rest. For an observer in this frame, the electron moves along a circular path within a fixed plane that is
perpendicular to the direction of the wave propagation, and according to
which the radiation spectrum must coincide with that of a synchrotron
radiation. This frame is characterized by the conditions \cite{BagGit2000},%
\begin{equation}
\boldsymbol{\kappa }=\mathbf{0}\,,\ \ p_{-}=mc\sqrt{1+\left( \frac{e}{mc^{2}}%
\right) ^{2}\left\langle \mathbf{A}^{2}\right\rangle }\,,  \label{pp2}
\end{equation}%
where $\left\langle \mathbf{A}^{2}\right\rangle $ is the averaged squared
vector potential, $\left\langle \mathbf{A}^{2}\right\rangle =\left( E_{0}/k_{%
\mathrm{w}}^{0}\right) ^{2}$. Imposing the conditions (\ref{pp2}) on Eq. (%
\ref{we}) we see that the energy in this frame takes the form%
\begin{eqnarray}
\overline{W}\left( \Delta \varphi \right) &=&\frac{e_{0}^{2}}{\pi c\omega _{%
\mathrm{w}}^{2}}\sum_{n=-\infty }^{+\infty }\int_{0}^{\infty }d\omega \omega
^{2}\int_{0}^{\pi }d\theta _{\gamma }\sin \theta _{\gamma }T_{n}\left( 
\overline{\eta },\Delta \varphi \right)  \notag \\
&\times &\left[ \frac{\left( n\omega _{\mathrm{w}}/\omega \right) ^{2}-\sin
^{2}\theta _{\gamma }}{\sin ^{2}\theta _{\gamma }}J_{n}^{2}\left( \overline{%
\mu }\right) +\frac{\xi ^{2}}{1+\xi ^{2}}J_{n}^{\prime 2}\left( \overline{%
\mu }\right) \right] \,,  \label{pp2b}
\end{eqnarray}%
where $\overline{\eta }=\omega /\omega _{\mathrm{w}}$, $\overline{\mu }=\xi
\omega \sin \theta _{\gamma }/\omega _{\mathrm{w}}\sqrt{1+\xi ^{2}}$. We
added a horizontal bar\ above the energy to distinguish it from the energy
given in Eq. (\ref{we}) as it refers to the energy emitted from the electron
in this frame. Similarly to the previous discussion, the energy (\ref{pp2b})
is finite because the radiation is generated within the interval $\Delta t$.
The maximum energy radiated from the electron is concentrated around the
resonance frequency $\overline{\omega }_{\mathrm{res}}=n\omega _{\mathrm{w}}$%
, with a width of $\Delta \overline{\omega }=4\pi /\Delta t$, and which has
the form,%
\begin{equation}
\overline{W}\left( \Delta \varphi \right) _{\max }=2\Delta tw_{\mathrm{Sch}%
}\,,  \label{pp5}
\end{equation}%
where $w_{\mathrm{Sch}}$ coincides with Schott's formula \cite{Schott} for
the energy rate radiated from an electron performing a circular motion,%
\begin{equation}
w_{\mathrm{Sch}}=\frac{\omega _{\mathrm{w}}^{2}e_{0}^{2}}{c}%
\sum_{n=1}^{\infty }n^{2}\int_{0}^{\pi }d\theta _{\gamma }\sin \theta
_{\gamma }\left[ \cot ^{2}\theta _{\gamma }J_{n}^{2}\left( \frac{n\xi \sin
\theta _{\gamma }}{\sqrt{1+\xi ^{2}}}\right) +\frac{\xi ^{2}}{1+\xi ^{2}}%
J_{n}^{\prime 2}\left( \frac{n\xi \sin \theta _{\gamma }}{\sqrt{1+\xi ^{2}}}%
\right) \right] \,.  \label{schott}
\end{equation}%
In this frame, the frequency of the plane wave $\omega _{\mathrm{w}}$ is
also the frequency of the electron's circular motion. This result was
also derived by Ritus in the context of QED \cite{Ritus85} through the
classical limit of the rate at which one photon is emitted from one electron
in a circularly polarized plane wave field.

It should be noted that Schott's formula (\ref{schott}) can also be
derived from the semiclassical energy rate,%
\begin{eqnarray}
\overline{w}\left( \Delta \varphi \right) &=&\frac{\partial }{\partial t}%
\overline{W}\left( \Delta \varphi \right) =\frac{e_{0}^{2}}{\pi c\omega _{%
\mathrm{w}}}\sum_{n=-\infty }^{+\infty }\int_{0}^{\infty }d\omega \omega
^{2}\int_{0}^{\pi }d\theta _{\gamma }\sin \theta _{\gamma }  \notag \\
&\times &\left[ \frac{\left( n\omega _{\mathrm{w}}/\omega \right) ^{2}-\sin
^{2}\theta _{\gamma }}{\sin ^{2}\theta _{\gamma }}J_{n}^{2}\left( \overline{%
\mu }\right) +\frac{\xi ^{2}}{1+\xi ^{2}}J_{n}^{\prime 2}\left( \overline{%
\mu }\right) \right] \frac{\sin \left[ \left( \overline{\eta }-n\right)
\Delta t\right] }{\left( \overline{\eta }-n\right) }\,,  \label{sr0.1}
\end{eqnarray}%
in the limit $\Delta t\rightarrow \infty $, $\lim_{\Delta t\rightarrow
\infty }\overline{w}\left( \Delta \varphi \right) =w_{\mathrm{Sch}}$\ due to
the identity (\ref{pp0}).

\subsection{Linearly-polarized external field\label{Sec2.2}}

In this section, we study the electromagnetic energy and energy rate emitted
from an electron in a linearly polarized plane wave field. The vector
potential of this field is a particular case of the one given in Eq. (\ref{A}%
) and can be chosen as follows%
\begin{equation}
A^{\mu }\left( \varphi \right) =a^{\mu }\sin \varphi \,,\ \ a^{\mu }=\left(
0,-\frac{cE_{0}}{\omega _{\mathrm{w}}},0,0\right) \,.  \label{lp1}
\end{equation}%
Using Eqs. (\ref{3.1}), (\ref{P}) and setting $\boldsymbol{\kappa }=\mathbf{0%
}$ for simplicity, we can easily show that the trajectory and momentum of an
electron in this field reads:%
\begin{eqnarray}
x\left( \varphi \right) &=&\ell _{x}\cos \varphi \,,\ \ y\left( \varphi
\right) =0\,,  \notag \\
z\left( \varphi \right) &=&\frac{\ell _{x}}{2\xi }\left( \frac{mc}{p_{-}}%
\right) \left( \lambda _{-}\varphi -\frac{\xi ^{2}}{4}\sin 2\varphi \right)
\,,  \notag \\
ct\left( \varphi \right) &=&\frac{\ell _{x}}{2\xi }\left( \frac{mc}{p_{-}}%
\right) \left( \lambda _{+}\varphi -\frac{\xi ^{2}}{4}\sin 2\varphi \right)
\,,  \label{lp2}
\end{eqnarray}%
and%
\begin{eqnarray}
P_{x}\left( \varphi \right) &=&-mc\xi \sin \varphi \,,\ \ P_{y}\left(
\varphi \right) =0\,,  \notag \\
P_{z}\left( \varphi \right) &=&q_{z}+\frac{p_{-}}{2}\left( \frac{mc}{p_{-}}%
\xi \right) ^{2}\sin ^{2}\varphi \,,  \notag \\
P^{0}\left( \varphi \right) &=&q^{0}+\frac{p_{-}}{2}\left( \frac{mc}{p_{-}}%
\xi \right) ^{2}\sin ^{2}\varphi \,.  \label{lp3}
\end{eqnarray}%
Here, $\xi $ is the classical nonlinearity parameter (\ref{radius})\footnote{%
For the linearly polarized plane wave field, this parameter is defined as $%
\xi =e_{0}\sqrt{-a_{\mu }a^{\mu }}/mc^{2}$ \cite{Ritus85}.}, $q_{z}$ and $%
q_{0}$ were defined previously in Eqs. (\ref{P3}), $\ell _{x}$ is the
amplitude of the electron motion in the $x$-direction and $\lambda _{\pm }$
are constants%
\begin{equation}
\ell _{x}=\frac{1}{k_{\mathrm{w}}^{0}}\left( \frac{mc}{p_{-}}\right) \xi =%
\frac{ce_{0}E_{0}}{\omega _{\mathrm{w}}^{2}p_{-}}\,,\ \ \lambda _{\pm }=1+%
\frac{\xi ^{2}}{2}\pm \left( \frac{p_{-}}{mc}\right) ^{2}\,.  \label{ell}
\end{equation}%
Note that the amplitude of motion in the $x$-direction $\ell _{x}$ formally
coincides with the radius of the electron's orbit in the plane perpendicular
to $\mathbf{n}_{\mathrm{w}}$ in the case of a circularly polarized plane
wave field (\ref{radius}).

By plugging the momenta (\ref{lp3}) into the current (\ref{de1c}) and
performing a change of variables $\nu ^{\prime }=\varphi ^{\prime }+\pi /2$
we can present the currents in the form%
\begin{eqnarray}
\tilde{j}^{0}\left( k;\Delta \varphi \right)  &=&-\frac{e^{iC^{\prime \prime
}}e_{0}}{4\pi ^{2}k_{\mathrm{w}}^{0}}\int_{\nu _{\mathrm{in}}}^{\nu }\left[ 
\frac{q^{0}}{p_{-}}+\frac{1}{2}\left( \frac{mc}{p_{-}}\xi \right) ^{2}\cos
^{2}\nu ^{\prime }\right] e^{i\psi \left( \nu ^{\prime }\right) }d\nu
^{\prime }\,,  \notag \\
\tilde{j}_{x}\left( k\mathbf{;}\Delta \varphi \right)  &=&-\frac{%
e^{iC^{\prime \prime }}e_{0}}{4\pi ^{2}k_{\mathrm{w}}^{0}}\left( \frac{mc}{%
p_{-}}\xi \right) \int_{\nu _{\mathrm{in}}}^{\nu }\cos \nu ^{\prime
}e^{i\psi \left( \nu ^{\prime }\right) }d\nu ^{\prime }\,,  \notag \\
\tilde{j}_{z}\left( k\mathbf{;}\Delta \varphi \right)  &=&-\frac{%
e^{iC^{\prime \prime }}e_{0}}{4\pi ^{2}k_{\mathrm{w}}^{0}}\int_{\nu _{%
\mathrm{in}}}^{\nu }\left[ \frac{q_{z}}{p_{-}}+\frac{1}{2}\left( \frac{mc}{%
p_{-}}\xi \right) ^{2}\cos ^{2}\nu ^{\prime }\right] e^{i\psi \left( \nu
^{\prime }\right) }d\nu ^{\prime }\,,  \label{lp6}
\end{eqnarray}%
where $C^{\prime \prime }$ is an unimportant constant phase, $\psi \left(
\nu ^{\prime }\right) =\sigma \nu ^{\prime }-\varrho \sin 2\nu ^{\prime
}-\zeta \sin \nu ^{\prime }$, and%
\begin{equation}
\sigma =\frac{\lambda _{+}-\lambda _{-}n_{z}}{2}\left( \frac{mc}{p_{-}}%
\right) ^{2}\frac{k^{0}}{k_{\mathrm{w}}^{0}}\,,\ \ \zeta =n_{x}\frac{mc}{%
p_{-}}\xi \frac{k^{0}}{k_{\mathrm{w}}^{0}}\,,\ \ \varrho =\frac{n_{z}-1}{8}%
\left( \frac{mc}{p_{-}}\xi \right) ^{2}\frac{k^{0}}{k_{\mathrm{w}}^{0}}\,.
\label{lp5}
\end{equation}%
The current $\tilde{j}_{y}\left( k\mathbf{;}\Delta \varphi \right) $ is
trivial due to the equations of motion (\ref{lp3}). Recall that $n_{x}=\sin
\theta _{\gamma }\cos \varphi _{\gamma }$ and $n_{z}=\cos \theta _{\gamma }$%
. Expanding the exponentials in terms of Bessel functions, as shown in the
first equation of (\ref{bessel}) and changing the summation index of one of
the Bessel functions, we can derive the following identity,%
\begin{equation}
e^{i\psi \left( \nu ^{\prime }\right) +il\nu ^{\prime }}=\sum_{n^{\prime
},n=-\infty }^{+\infty }J_{n^{\prime }}\left( \varrho \right)
J_{n-2n^{\prime }+l}\left( \zeta \right) e^{i\left( \sigma -n\right) \nu
^{\prime }}\,,  \label{lp9}
\end{equation}%
which can then be used to express the currents (\ref{lp6}) in the form:%
\begin{eqnarray}
\tilde{j}^{0}\left( k;\Delta \varphi \right)  &=&-\frac{e^{iC^{\prime \prime
}}e_{0}}{4\pi ^{2}k_{\mathrm{w}}^{0}}\sum_{n=-\infty }^{+\infty }\left[ 
\frac{q^{0}}{p_{-}}\mathcal{A}_{n}^{\left( 0\right) }\left( \varrho ,\zeta
\right) +\frac{1}{2}\left( \frac{mc}{p_{-}}\xi \right) ^{2}\mathcal{A}%
_{n}^{\left( 2\right) }\left( \varrho ,\zeta \right) \right] \int_{\nu _{%
\mathrm{in}}}^{\nu }e^{i\left( \sigma -n\right) \nu ^{\prime }}d\nu ^{\prime
}\,,  \notag \\
\tilde{j}_{x}\left( k\mathbf{;}\Delta \varphi \right)  &=&-\frac{%
e^{iC^{\prime \prime }}e_{0}}{4\pi ^{2}k_{\mathrm{w}}^{0}}\sum_{n=-\infty
}^{+\infty }\left( \frac{mc}{p_{-}}\xi \right) \mathcal{A}_{n}^{\left(
1\right) }\left( \varrho ,\zeta \right) \int_{\nu _{\mathrm{in}}}^{\nu
}e^{i\left( \sigma -n\right) \nu ^{\prime }}d\nu ^{\prime }\,,  \notag \\
\tilde{j}_{z}\left( k\mathbf{;}\Delta \varphi \right)  &=&-\frac{%
e^{iC^{\prime \prime }}e_{0}}{4\pi ^{2}k_{\mathrm{w}}^{0}}\sum_{n=-\infty
}^{+\infty }\left[ \frac{q_{z}}{p_{-}}\mathcal{A}_{n}^{\left( 0\right)
}\left( \varrho ,\zeta \right) +\frac{1}{2}\left( \frac{mc}{p_{-}}\xi
\right) ^{2}\mathcal{A}_{n}^{\left( 2\right) }\left( \varrho ,\zeta \right) %
\right] \int_{\nu _{\mathrm{in}}}^{\nu }e^{i\left( \sigma -n\right) \nu
^{\prime }}d\nu ^{\prime }\,,  \label{lp9b}
\end{eqnarray}%
where%
\begin{eqnarray*}
\mathcal{A}_{n}^{\left( 0\right) }\left( \varrho ,\zeta \right) 
&=&\sum_{n^{\prime }=-\infty }^{+\infty }J_{n^{\prime }}\left( \varrho
\right) J_{n-2n^{\prime }}\left( \zeta \right) \,, \\
\mathcal{A}_{n}^{\left( 1\right) }\left( \varrho ,\zeta \right)  &=&\frac{%
\mathcal{A}_{n+1}^{\left( 0\right) }\left( \varrho ,\zeta \right) +\mathcal{A%
}_{n-1}^{\left( 0\right) }\left( \varrho ,\zeta \right) }{2}\,, \\
\mathcal{A}_{n}^{\left( 2\right) }\left( \varrho ,\zeta \right)  &=&\frac{%
\mathcal{A}_{n+2}^{\left( 0\right) }\left( \varrho ,\zeta \right) +2\mathcal{%
A}_{n}^{\left( 0\right) }\left( \varrho ,\zeta \right) +\mathcal{A}%
_{n-2}^{\left( 0\right) }\left( \varrho ,\zeta \right) }{4}\,.
\end{eqnarray*}

Finally, Eq. (\ref{de3}) admits the general structure:%
\begin{eqnarray}
W\left( \Delta \varphi \right) &=&\frac{1}{2}\left( \frac{e_{0}mc^{2}}{\pi
\omega _{\mathrm{w}}p_{-}}\right) ^{2}\sum_{n=-\infty }^{+\infty }\int
d\Omega \int_{0}^{\infty }d\omega \omega ^{2}\left\{ -\left( \mathcal{A}%
_{n}^{\left( 0\right) }\left( \varrho ,\zeta \right) \right) ^{2}\right. 
\notag \\
&+&\left. \xi ^{2}\left[ \left( \mathcal{A}_{n}^{\left( 1\right) }\left(
\varrho ,\zeta \right) \right) ^{2}-\mathcal{A}_{n}^{\left( 0\right) }\left(
\varrho ,\zeta \right) \mathcal{A}_{n}^{\left( 2\right) }\left( \varrho
,\zeta \right) \right] \right\} T_{n}\left( \sigma ,\Delta \varphi \right) 
\mathbf{\,.}  \label{lp12}
\end{eqnarray}%
where $T_{n}\left( \sigma ,\Delta \varphi \right) $ is defined in Eq. (\ref%
{we2}). This expression is analogous to the electromagnetic energy emitted
from an electron in a circularly polarized plane wave field (\ref{we}) and
represents the main result of this section. It describes the semiclassical
electromagnetic energy radiated from an electron in a linearly polarized
plane wave field within the phase interval $\Delta \varphi $. Similarly to
the preceding case, the energy (\ref{lp12}) is finite owing to the presence
of the oscillatory function $T_{n}\left( \sigma ,\Delta \varphi \right) $.
The latter depends on the phase interval $\Delta \varphi =\varphi -\varphi _{%
\mathrm{in}}$, which is linked to the quantum radiation transition interval $%
\Delta t=t-t_{\mathrm{in}}$ as discussed in Sec. \ref{Sec2.0}. As stated
before, this function is undefined in the classical limit $\Delta \varphi
\rightarrow \infty $ but it is sharply peaked at the resonance frequency $%
\omega _{\mathrm{res}}$, which in this case is given by%
\begin{equation}
\omega _{\mathrm{res}}=n\omega _{\mathrm{r}}\,,\ \ \omega _{\mathrm{r}%
}=\omega _{\mathrm{w}}\frac{2\left( p_{-}/mc\right) ^{2}}{\lambda
_{+}-\lambda _{-}\cos \theta _{\gamma }}\,,  \label{lp14}
\end{equation}%
with a characteristic width inversely proportional to the phase interval $%
\Delta \phi =\phi -\phi _{\mathrm{in}}$,%
\begin{equation}
\Delta \omega =\frac{\left( p_{-}/mc\right) ^{2}}{\lambda _{+}-\lambda
_{-}\cos \theta _{\gamma }}\frac{8\pi }{\Delta \phi }\,.  \label{lp15}
\end{equation}%
In contrast to the\ case of a circularly polarized plane wave field, the
width (\ref{lp15}) cannot be expressed in terms of the quantum transition
interval $\Delta t$ because the electron's trajectory (\ref{lp2}) and its
momentum (\ref{lp3}) cannot be defined as functions of time. As a result,
the maximum energy $W\left( \Delta \varphi \right) _{\max }$ emitted from
the electron in this case is proportional to $\Delta \phi $:%
\begin{eqnarray}
W\left( \Delta \varphi \right) _{\max } &\approx &\Delta \phi \frac{\left(
e_{0}c\omega _{\mathrm{w}}\right) ^{2}}{2\pi }\left( \frac{2p_{-}}{mc}%
\right) ^{4}\sum_{n=1}^{\infty }n^{2}\int \frac{d\Omega }{\left( \lambda
_{+}-\lambda _{-}\cos \theta _{\gamma }\right) ^{3}}  \notag \\
&\times &\left\{ -\left( \mathcal{A}_{n}^{\left( 0\right) }\left( \omega _{%
\mathrm{res}}\right) \right) ^{2}+\xi ^{2}\left[ \left( \mathcal{A}%
_{n}^{\left( 1\right) }\left( \omega _{\mathrm{res}}\right) \right) ^{2}-%
\mathcal{A}_{n}^{\left( 0\right) }\left( \omega _{\mathrm{res}}\right) 
\mathcal{A}_{n}^{\left( 2\right) }\left( \omega _{\mathrm{res}}\right) %
\right] \right\} \,,  \label{lp16}
\end{eqnarray}%
where $\mathcal{A}_{n}^{\left( j\right) }\left( \omega _{\mathrm{res}%
}\right) =\mathcal{A}_{n}^{\left( j\right) }\left( \varrho _{\mathrm{res}%
},\zeta _{\mathrm{res}}\right) $ and%
\begin{equation*}
\varrho _{\mathrm{res}}=n\frac{\xi ^{2}}{4}\frac{\cos \theta _{\gamma }-1}{%
\lambda _{+}-\lambda _{-}\cos \theta _{\gamma }}\,,\ \ \zeta _{\mathrm{res}%
}=2n\left( \frac{p_{-}}{mc}\xi \right) \frac{\sin \theta _{\gamma }\cos
\varphi _{\gamma }}{\lambda _{+}-\lambda _{-}\cos \theta _{\gamma }}\,.
\end{equation*}%
To estimate the maximum energy (\ref{lp16}), we took into account that
radiation is generated within a sufficiently large phase interval $\Delta
\phi $. Under this condition, we approximated the oscillatory function by
its leading term (\ref{we5b}) and replaced the integral over $\omega $ with
its main contribution, which comes from a narrow bandwidth centered at
the resonance frequency (\ref{lp14}). The summation over negative $n$ does
not contribute to (\ref{lp16}) as the frequency $\omega _{\mathrm{r}}$ (\ref%
{lp14}) is positive and the integration interval in (\ref{lp12}) is positive.

The maximum radiation spectrum (\ref{lp16}) is directly proportional to the
phase interval $\Delta \phi $ rather than the time interval $\Delta t$. As
discussed above, this is because the electron's trajectory in the
configuration space cannot be parameterized by the laboratory time.
Therefore, for this external field, the phase $\phi =\left( n_{\mathrm{w}%
}x\right) /c$ plays the role of time, and we find it appropriate to identify the
right-hand side of Eq. (\ref{lp16}) divided by $\Delta \phi $ as the maximum
energy rate emitted from the electron in this field,%
\begin{equation}
w_{\max }=\frac{W\left( \Delta \varphi \right) _{\max }}{\Delta \phi }\,.
\label{lp17}
\end{equation}%
Additionally, the spectrum (\ref{lp16}) diverges if the electron interacts
with the field over an infinite phase interval $\Delta \phi $. This type of
divergence was previously discussed by Ritus in the context of the classical
theory; see Ref. \cite{Ritus85}. In his work, Ritus heuristically related
the phase interval during which radiation is produced $\Delta\varphi$ with time and derived an
expression for the energy rate emitted from the electron in a linearly
polarized plane wave field. Aside from differences in the signs of the
parameters $\varrho $ and $\zeta $ (\ref{lp5}) (which can be traced back to
the choice of the potential (\ref{lp1}) and does not affect the spectrum (%
\ref{lp16})), and numerical constants related to Ritus's connection between
phase and time, our result (\ref{lp17}) coincides with his. The maximum rate
(\ref{lp17}) can be compared with the semiclassical energy rate $w\left(
\Delta \varphi \right) $, which we define as the derivative of the energy (%
\ref{lp12}) with respect to the phase $\phi $:%
\begin{eqnarray}
w\left( \Delta \varphi \right) &=&\frac{\partial }{\partial \phi }W\left(
\Delta \varphi \right) =\frac{1}{2\omega _{\mathrm{w}}}\left( \frac{%
e_{0}mc^{2}}{\pi p_{-}}\right) ^{2}\sum_{n=-\infty }^{+\infty }\int d\Omega
\int_{0}^{\infty }d\omega \omega ^{2}\left\{ -\left( \mathcal{A}_{n}^{\left(
0\right) }\left( \varrho ,\zeta \right) \right) ^{2}\right.  \notag \\
&+&\left. \xi ^{2}\left[ \left( \mathcal{A}_{n}^{\left( 1\right) }\left(
\varrho ,\zeta \right) \right) ^{2}-\mathcal{A}_{n}^{\left( 0\right) }\left(
\varrho ,\zeta \right) \mathcal{A}_{n}^{\left( 2\right) }\left( \varrho
,\zeta \right) \right] \right\} \frac{\sin \left[ \left( \sigma -n\right)
\Delta \varphi \right] }{\left( \sigma -n\right) }\,.  \label{lp18}
\end{eqnarray}%
In the classical limit $\Delta \varphi \rightarrow \infty $, we can use the
identify (\ref{pp0}) to show that the classical energy rate $w_{\mathrm{cl}}$
is half of the maximum estimate (\ref{lp17}),%
\begin{equation}
w_{\mathrm{cl}}=\lim_{\Delta \varphi \rightarrow \infty }w\left( \Delta
\varphi \right) =\frac{w_{\max }}{2}\,.  \label{lp19}
\end{equation}%
The same relation was obtained in the case of a circularly polarized plane
wave field, see Eq. (\ref{we7}).

To conclude this section, we present the energy and the energy rate emitted
from the electron in the frame where the electron is on average at rest.
This frame is characterized by the conditions (\ref{pp2}) with $\left\langle 
\mathbf{A}^{2}\right\rangle =\left( E_{0}/\sqrt{2}k_{\mathrm{w}}^{0}\right)
^{2}$. In this frame, the electron performs a periodic motion in the shape
of a figure-8 in the $xz$-plane and the semiclassical energy spectrum (\ref%
{lp12}) takes the form%
\begin{eqnarray}
\overline{W}\left( \Delta \varphi \right) &=&\frac{\left( e_{0}c\right) ^{2}%
}{2\pi ^{2}\omega _{\mathrm{w}}^{2}\left( 1+\xi ^{2}/2\right) }%
\sum_{n=-\infty }^{+\infty }\int d\Omega \int_{0}^{\infty }d\omega \omega
^{2}\left\{ -\left( \mathcal{A}_{n}^{\left( 0\right) }\left( \overline{%
\varrho },\overline{\zeta }\right) \right) ^{2}\right.  \notag \\
&+&\left. \xi ^{2}\left[ \left( \mathcal{A}_{n}^{\left( 1\right) }\left( 
\overline{\varrho },\overline{\zeta }\right) \right) ^{2}-\mathcal{A}%
_{n}^{\left( 0\right) }\left( \overline{\varrho },\overline{\zeta }\right) 
\mathcal{A}_{n}^{\left( 2\right) }\left( \overline{\varrho },\overline{\zeta 
}\right) \right] \right\} T_{n}\left( \overline{\sigma },\Delta \varphi
\right) \mathbf{\,,}  \label{lp20}
\end{eqnarray}%
where%
\begin{equation}
\overline{\sigma }=\frac{\omega }{\omega _{\mathrm{w}}}\,,\ \ \overline{%
\varrho }=n\frac{\xi ^{2}}{8}\frac{\cos \theta _{\gamma }-1}{1+\xi ^{2}/2}%
\,,\ \ \overline{\zeta }=\frac{n\xi }{\sqrt{1+\xi ^{2}/2}}\sin \theta
_{\gamma }\cos \varphi _{\gamma }\,.  \label{lp21}
\end{equation}%
For an observer in this frame, the maximum energy radiated from the electron
is near the resonance frequency $\overline{\omega }_{\mathrm{res}}=\omega _{%
\mathrm{w}}$, with a width approximately given by $\Delta \overline{\omega }%
=4\pi /\Delta \phi $, and has the form%
\begin{equation}
\overline{W}\left( \Delta \varphi \right) =8\pi \Delta \phi \overline{w}_{%
\mathrm{cl}}\,,  \label{lp22}
\end{equation}%
where $\overline{w}_{\mathrm{cl}}$ is the classical energy rate obtained by
Nikishov and Ritus in \cite{NikRit64},%
\begin{eqnarray}
\overline{w}_{\mathrm{cl}} &=&\frac{\left( e_{0}c\omega _{\mathrm{w}}\right)
^{2}}{8\pi ^{2}\left( 1+\xi ^{2}/2\right) }\sum_{n=1}^{\infty }n^{2}\int
d\Omega \left\{ -\left( \mathcal{A}_{n}^{\left( 0\right) }\left( \overline{%
\varrho },\overline{\zeta }\right) \right) ^{2}\right.  \notag \\
&+&\left. \xi ^{2}\left[ \left( \mathcal{A}_{n}^{\left( 1\right) }\left( 
\overline{\varrho },\overline{\zeta }\right) \right) ^{2}-\mathcal{A}%
_{n}^{\left( 0\right) }\left( \overline{\varrho },\overline{\zeta }\right) 
\mathcal{A}_{n}^{\left( 2\right) }\left( \overline{\varrho },\overline{\zeta 
}\right) \right] \right\} \,.  \label{lp23}
\end{eqnarray}

\section{Concluding remarks}

We discussed the electromagnetic energy and energy rate spectra radiated
from an electron in monochromatic plane wave fields. The study is based on a
semiclassical formulation \cite{BagGiSF20,ShiLeBG21,TGJB}, where currents
are treated as classical quantities while the electromagnetic field is
considered quantum. In this framework, the energy spectrum is derived from
the transition probability of the Schr\"{o}dinger state to evolve from an
initial state without photons at time $t_{\mathrm{in}}$ to a final state
with photons at time $t$. As a result, the transition time interval between
quantum states $\Delta t$ is introduced at a fundamental level and is
reflected in the Fourier transform of the current density. The resulting
energy spectrum carries this time interval and is free of divergences
associated with the duration over which the particle is accelerated by the
external field. This feature is not present in classical theory.

We considered two external fields: circularly polarized and linearly
polarized plane wave fields. In both instances, the semiclassical energy
features an oscillatory function that regulates the spectra and favors
electromagnetic radiation around a characteristic resonance frequency $\omega_{\mathrm{res}}$. In
the first instance, the maximum energy spectrum radiated from the particle is
linearly proportional to the transition interval. Such a time dependence not
only allowed us to estimate the maximum energy rate clearly but also enabled
us to isolate the information associated with the time during which the
particle interacts with the external field. Since the particle-external
field interaction time is contained within the quantum transition interval $%
\Delta t$, it becomes evident that the energy radiated by the particle
diverges if it interacts indefinitely with the external field. While this
argument is intuitive, to our knowledge, this specific type of divergence
has not yet been explicitly discussed in the literature. One possible
explanation for this is that classical electromagnetic energy is derived
from the energy rate through an integral over an infinite time interval. If
the rate is finite and if the particle interacts with the external field
indefinitely, then the energy spectrum is intrinsically divergent, which creates difficulties in defining the total radiated electromagnetic energy. 

When calculating the semiclassical energy spectrum radiated from an electron in a linearly polarized plane wave field, we observed that the maximum
spectrum is linearly proportional to the phase interval $\Delta \phi $ over
which the radiation occurs, rather than to the time interval $\Delta t$.
This is a consequence of the fact that the electron's motion cannot be
parameterized by the laboratory time. Nevertheless, because the phase of the
wave naturally serves a \textquotedblleft time\textquotedblright\ for this
field, we identified the corresponding energy rate spectrum from the energy
spectrum and realized that our result is compatible with Ritus's \cite%
{Ritus85} in the classical limit. By specializing our results to a special
reference frame, where the electron is at rest on average, we reproduced results compatible with those obtained earlier in the context of
classical electrodynamics. Specifically, we derived Schott's formula \cite%
{Schott} in the case of a circularly polarized field and Nikishov-Ritus's
formula \cite{NikRit64} in the case of a linearly polarized field. It is noteworthy that Nikishov and Ritus obtained the classical energy rate
spectra within the framework of QED with external fields, specifically from
transition probabilities corresponding to the emission of one
photon from an electron in plane wave fields through a classical limit. In
the semiclassical formulation discussed in this work, we compute the energy
and energy rate spectra through transition probabilities from the initial
state without photons to the final state containing an infinite
number of photons. This corresponds to Eqs. (26) - (29) in our previous work 
\cite{TGJB} and Eq. (\ref{de1}) above. Thus, in a general sense, the
compatibility between the semiclassical formulation and classical theory is
consistent with Heitler's interpretation \cite{Heitl36}, which states that
the classical limit of quantities calculated in the quantum theory of
radiation is achieved when the number of photons is sufficiently large%
\footnote{%
See also the textbook \cite{Sakurai} for a discussion.}. On the other
hand, we note that the correspondence between QED and classical
electrodynamics through a classical limit is consistent with Akhiezer's and
Berestetskii's interpretation \cite{AkhBe65}, which states that Maxwell's
equations for the electromagnetic field can be understood as the Schr\"{o}%
dinger equation for a single photon\footnote{%
The absence of $\hslash $ in the Schr\"{o}dinger equation is due to the
triviality of photon's mass.}. From this perspective, electromagnetic
quantities calculated using Maxwell's fields (such as electromagnetic
radiation) are related to properties of a single photon. We do not
advocate for one interpretation over the other, but acknowledge the
coexistence of both perspectives. Regarding the compatibility between the
semiclassical formulation and QED, we believe that it stems from the nature
of the current density. This subject will be a topic of a future study.

To conclude this work, we emphasize that the semiclassical formulation
offers a consistent framework for calculating the energy and energy rate
spectra emitted from current distributions accelerated by external fields.
While the energy spectrum is classical in nature--meaning it does not
contain parameters inherent to quantum theory, such as the fine structure
constant or the quantum nonlinearity parameter (\ref{chie})--its origin is purely quantum
as it is derived from a transition probability. We do not repeat here the
derivation leading to Eq. (\ref{de1}) as it is detailed in our previous work 
\cite{TGJB}. Instead, we employed the main results to study
radiation in plane wave fields, which are the basis for more realistic
fields that can be reproduced in laboratory settings and used to investigate
important phenomena arising from the interaction between light and matter
under extreme conditions. Finally, it is worth mentioning that the
semiclassical formulation does not account for effects related to the spin
of the current density. However, it does allow for including radiation-reaction
effects, provided that solutions to relativistic equations with
radiation-reaction terms, such as the Landau-Lifshitz equation, exist.

\section*{Acknowledgments}

T. C. A. acknowledges financial support from XJTLU Research Development
Funding, Award No. RDF-21-02-056. D. M. G. thanks FAPESP (Grant No. 21/10128-0) and CNPq for permanent support.


\begin{thebibliography}{99}
\bibitem{Aharonian00} F. A. Aharonian, \emph{TeV gamma rays from BL Lac
objects due to synchrotron radiation of extremely high energy protons}, New
Astronomy \textbf{5}, 377 (2000).

\bibitem{GaeSla06} P. Gaensler and P. O. Slane, \emph{The Evolution and
Structure of Pulsar Wind Nebulae} Ann. Rev. Astron. Astrophys. \textbf{44},
17 (2006).

\bibitem{Wataru-etal17} W. Ishizaki, S. J. Tanaka, K. Asano, and T.
Terasawa, \emph{Broadband Photon Spectrum and its Radial Profile of Pulsar
Wind Nebulae}, Astrophys. J. \textbf{838} 142 (2017).

\bibitem{Reynolds-etal17} S. Reynolds, G. G. Pavlov, O. Kargaltsev \textit{%
et al.}, \emph{Pulsar-Wind Nebulae and Magnetar Outflows: Observations at
Radio, X-Ray, and Gamma-Ray Wavelengths} Space Sci Rev \textbf{207}, 175
(2017).

\bibitem{Padovani21} M. Padovani, A. Bracco, V. Jeli\'{c}, D. Galli, and E.
Bellomi, \emph{Spectral index of synchrotron emission: insights from the
diffuse and magnetised interstellar medium}, Astron. \& Astrophys. \textbf{%
651}, A116 (2021).

\bibitem{HarLai06} A. K. Harding and D. Lai, \emph{Physics of strongly
magnetized neutron stars}, Rep. Prog. Phys. \textbf{69}, 2631 (2006).

\bibitem{KelProAha15} S. R. Kelner, A. Yu. Prosekin, and F. A. Aharonian, \emph{Synchro-curvature radiation of charged particles in the strong curved magnetic fields}, Astron. J. \textbf{149}, 33 (2015).

\bibitem{BerPulVol09} E. G. Berezhko, G. P\"{u}hlhofer, and H. J. V\"{o}lk, \emph{Theory of cosmic ray and $\gamma$-ray production in the supernova remnant RX J0852.0-4622}, Astron. $\&$ Astrophys. \textbf{505}, 641 (2009).

\bibitem{StrOrlJaf11} A. W. Strong, E. Orlando, and T. R. Jaffe, \emph{The
interstellar cosmic-ray electron spectrum from synchrotron radiation and
direct measurements}, Astron. \& Astrophys. \textbf{534}, A54 (2011).

\bibitem{KotAllLem11} K. Kotera, D. Allard, and M. Lemoine, \emph{%
Detectability of ultrahigh energy cosmic-ray signatures in gamma-rays},
Astron. \& Astrophys. \textbf{527}, A54 (2011).

\bibitem{Kenta18} K. Hotokezaka, K. Kiuchi, M. Shibata \textit{et al.}, \emph{Synchrotron Radiation from the Fast Tail of Dynamical Ejecta of Neutron Star Mergers}, Astrophys. J. \textbf{867}, 95 (2018).

\bibitem{ESRF} European Synchrotron Radiation Facility -- ESRF,
https://www.esrf.fr/

\bibitem{Patrick-etal24} B. Patrick \textit{et al.}, \emph{X-ray science
using the ESRF--extremely brilliant source} Eur. Phys. J. Plus \textbf{139},
928 (2024).

\bibitem{Schotta} G. A. Schott, \emph{On the Radiation from Moving Systems
of Electrons and on the Spectrum of Canal Rays} Phil. Mag. \textbf{13}, 657
(1907).

\bibitem{Schottb} G. A. Schott, \emph{\"{U}ber die Strahlung von
Elektronengruppen} Ann. Physik. \textbf{329}, 635 (1907).

\bibitem{Schott} G. A. Schott, \emph{Electromagnetic Radiation} (Cambrige
University Press, Cambrige 1912).

\bibitem{Schwinger49} J. Schwinger, \emph{On the Classical Radiation of
Accelerated Electrons}, Phys. Rev. \textbf{75}, 1912 (1949).

\bibitem{SokTe68} A. A. Sokolov, I. M. Ternov, \emph{Synchrotron Radiation}
(Academic Verlag, Berlin 1968); \emph{Radiation from relativistic electrons}{%
\ (American Institute of Physics, New York 1986).}

\bibitem{Bordovitsyn99} V. A. Bordovitsyn, \emph{Synchrotron Radiation
Theory and Its Development} (World Scientific, Singapore, 1999).

\bibitem{Wiedemann} H. Wiedemann, \emph{Synchrotron Radiation} (Springer,
Berlin, 2003).

\bibitem{Schwbook} J. Schwinger, L. L. Deraad Jr., K. Milton, and W.-Y.
Tsai, \emph{Classical Electrodynamics} (Westview Press Inc, 1st edition,
1998).

\bibitem{Jacks99} J. D. Jackson, \emph{Classical Electrodynamics}, (John
Wiley \& Sons, New York, 1999).

\bibitem{ItzZub} C. Itzykson, J.-B. Zuber, \emph{Quantum Field Theory}
(McGraw-Hill, New York 1980).

\bibitem{Reiss62} H. R. Reiss, \emph{Absorption of Light by Light}, J. Math.
Phys. \textbf{3}, 59, (1962).

\bibitem{Goldman64} I. I. Goldman, \emph{Intensity Effects in Compton
Scattering}, Sov. Phys. JETP \textbf{19}, 954, (1964).

\bibitem{BroKib64} L. S. Brown and T. W. B. Kibble, \emph{Interaction of
Intense Laser Beams with Electrons}, Phys. Rev. \textbf{133}, A705, (1964).

\bibitem{NikRit64} A. I. Nikishov and V. I. Ritus, \emph{Quantum Processes
in the Field of a Plane Electromagnetic Wave and in a Constant Field I},
Sov. Phys. JETP \textbf{19}, 529 (1964) [Zh. Eksp. Teor. Fiz. \textbf{46},
776 (1964)].

\bibitem{Volkov} D. M. Wolkow, \emph{\"{U}ber eine Klasse von L\"{o}sungen
der Diracschen Gleichung}, Z. Physik \textbf{94}, 250 (1935).

\bibitem{Ritus85} V. I. Ritus, \emph{Quantum effects of the interaction of
elementary particles with an intense electromagnetic field}, J. Russ. Laser
Res. \textbf{6}, 497 (1985).

\bibitem{Furry51} W. H. Furry, \emph{On Bound States and Scattering in
Positron Theory}, Phys. Rev. \textbf{81}, 115 (1951).\emph{\ }

\bibitem{XFEL} https://www.xfel.eu/

\bibitem{ELI} https://www.eli-laser.eu/

\bibitem{DESY} https://photon-science.desy.de/

\bibitem{LINAC} https://lcls.slac.stanford.edu/

\bibitem{HeiSeiKam10} T. Heinzl, D. Seipt, and B. K\"{a}mpfer, \emph{%
Beam-shape effects in nonlinear Compton and Thomson scattering}, Phys. Rev.
A \textbf{81}, 022125 (2010).

\bibitem{KraKam12} K. Krajewska and J. Z. Kami\'{n}ski, \emph{Compton
process in intense short laser pulses}, Phys. Rev. A \textbf{85}, 062102
(2012).

\bibitem{ShePia12} O. Har-Shemesh and A. Di Piazza, \emph{Peak intensity
measurement of relativistic lasers via nonlinear Thomson scattering}, Opt.
Lett. \textbf{37}, 1352 (2012).

\bibitem{Piazza18} A. Di Piazza, \emph{Analytical infrared limit of
nonlinear Thomson scattering including radiation reaction}, Phys. Lett. B, 
\textbf{782}, 559 (2018).

\bibitem{PiaAud21} A. Di Piazza and G. Audagnotto, \emph{Analytical spectrum
of nonlinear Thomson scattering including radiation reaction}, Phys. Rev. D 
\textbf{104}, 016007 (2021).

\bibitem{PiaTamMeuKei18} A. Di Piazza, M. Tamburini, S. Meuren, and C. H.
Keitel, \emph{Implementing nonlinear Compton scattering beyond the local
constant field approximation}, Phys. Rev. A \textbf{98}, 012134 (2018).

\bibitem{IldKinSei19} A. Ilderton, B. King, D. Seipt, \emph{An Extended
Locally Constant Field Approximation for Nonlinear Compton Scattering},
Phys. Rev. A \textbf{99}, 042121 (2019).

\bibitem{AngMacPia16} A. Angioi, F. Mackenroth, and A. Di Piazza, \emph{%
Nonlinear single Compton scattering of an electron wave packet}, Phys. Rev.
A \textbf{93}, 052102 (2016).

\bibitem{KinTan20} B. King and S. Tang, \emph{Nonlinear Compton Scattering
of Polarised Photons in Plane-Wave Backgrounds}, Phys. Rev. A \textbf{102},
022809 (2020).

\bibitem{IldKinTan20} A. Ilderton, B. King, and S. Tang, \emph{Toward the
observation of interference effects in nonlinear Compton scattering}, Phys.
Lett. B \textbf{804}, 135410 (2020).

\bibitem{Bula-etal96} C. Bula, K. T. MacDonald, E. J. Prebys \textit{et al.}%
, \emph{Observation of Nonlinear Effects in Compton Scattering}, Phys. Rev.
Lett. \textbf{76}, 3116 (1996).

\bibitem{JanMul13} M. J. A. Jansen and C. M\"{u}ller, \emph{Strongly
enhanced pair production in combined high- and low-frequency laser fields},
Phys. Rev. A \textbf{88}, 052125 (2013).

\bibitem{KraKa14} K Krajewska and J Z Kami\'{n}ski, \emph{Breit-Wheeler pair
creation by finite laser pulses}, J. Phys.: Conf. Ser. \textbf{497}, 012016
(2014).

\bibitem{WuXue14} Y.-B. Wu and S.-S. Xue, \emph{Nonlinear Breit-Wheeler
process in the collision of a photon with two plane waves}, Phys. Rev. D 
\textbf{90}, 013009 (2014).

\bibitem{MeuHatKeiPia15} S. Meuren, K. Z. Hatsagortsyan, C. H. Keitel, and
A. Di Piazza, \emph{Polarization-operator approach to pair creation in short
laser pulses}, Phys. Rev. D \textbf{91}, 013009 (2015).

\bibitem{Otto-etal16} A. Otto, T. Nousch, D. Seipt \textit{et al.}, \emph{%
Pair production by Schwinger and Breit--Wheeler processes in bi-frequent
fields}, J. Plasma Phys. \textbf{82}, 655820301 (2016).

\bibitem{DiPiazza16} A. Di Piazza, \emph{Nonlinear Breit-Wheeler Pair
Production in a Tightly Focused Laser Beam}, Phys. Rev. Lett. \textbf{117},
213201 (2016).

\bibitem{MeuKeiPia16} S. Meuren, C. H. Keitel, and A. Di Piazza, \emph{%
Semiclassical Picture for Electron-Positron Photoproduction in Strong Laser
Fields}, Phys. Rev. D \textbf{93}, 085028 (2016).

\bibitem{JanMul16} M. J. A. Jansen and C. M\"{u}ller, \emph{Strong-field
Breit-Wheeler pair production in short laser pulses: Identifying multiphoton
interference and carrier-envelope-phase effects}, Phys. Rev. D \textbf{93},
053011 (2016).

\bibitem{JanMul17} M. J. A. Jansen and C. M\"{u}ller, \emph{Strong-field
Breit--Wheeler pair production in two consecutive laser pulses with variable
time delay}, Phys. Lett. B \textbf{766}, 71 (2017).

\bibitem{GolChaMul22} A. Golub, S. Villalba-Ch\'{a}vez, and C. M\"{u}ller, 
\emph{Nonlinear Breit-Wheeler pair production in collisions of
bremsstrahlung }$\gamma $\emph{\ quanta and a tightly focused laser pulse},
Phys. Rev. D \textbf{105}, 116016 (2022).

\bibitem{MahChaMul23} N. Mahlin, S. Villalba-Ch\'{a}vez, and C. M\"{u}ller, 
\emph{Dynamically assisted nonlinear Breit-Wheeler pair production in
bichromatic laser fields of circular polarization}, Phys. Rev. D \textbf{108}%
, 096023 (2023).

\bibitem{HarIldKin15} C. N. Harvey, A. Ilderton, and B. King, \emph{Testing
numerical implementations of strong-field electrodynamics}, Phys. Rev. A 
\textbf{91}, 013822 (2015).

\bibitem{BlaSeiBulMar18} T. G. Blackburn, D. Seipt, S. S. Bulanov, and M.
Marklund, \emph{Benchmarking semiclassical approaches to strong-field QED:
Nonlinear Compton scattering in intense laser pulses}, Phys. Plasmas \textbf{%
25}, 083108 (2018).

\bibitem{BlaSeiBulMar20} T. G. Blackburn, D. Seipt, S. S. Bulanov, and M.
Marklund, \emph{Radiation beaming in the quantum regime}, Phys. Rev. A 
\textbf{101}, 012505 (2020).

\bibitem{MouTajBul06} G. A. Mourou, T. Tajima, and S. V. Bulanov, \emph{%
Optics in the relativistic regime}, Rev. Mod. Phys. \textbf{78}, 309 (2006).

\bibitem{MarShu06} M.\thinspace Marklund and P. K.\thinspace Shukla, \emph{%
Nonlinear collective effects in photon-photon and photon-plasma interactions}%
,\ Rev. Mod. Phys.\thinspace \textbf{78}, 591 (2006).

\bibitem{EhlKraKam09} F. Ehlotzky, K. Krajewska, and J. Z. Kami\'{n}ski, 
\emph{Fundamental processes of quantum electrodynamics in laser fields of
relativistic power}, Rep. Prog. Phys. 72 046401 (2009).

\bibitem{Bulanov-etal11} S. V.\thinspace Bulanov, T. Z.\thinspace Esirkepov,
Y.\thinspace Hayashi \textit{et al.} \emph{On the design of experiments for
the study of extreme field limits in the interaction of laser with
ultrarelativistic electron beam},\ Nucl. Instrum. Methods Phys. Res.
A\thinspace \textbf{660}, 31 (2011).

\bibitem{PiaMulHatKei12} A. DiPiazza, C.\thinspace M\"{u}ller, K.
Z.\thinspace Hatsagortsyan, and C. H.\thinspace Keitel, \emph{Extremely
high-intensity laser interactions with fundamental quantum systems},\ Rev.
Mod. Phys.\thinspace \textbf{84}, 1177 (2012).

\bibitem{Blackburn20} T. G. Blackburn, \emph{Radiation reaction in
electron--beam interactions with high-intensity lasers}, Rev. Mod. Plasma
Phys \textbf{4}, 5 (2020).

\bibitem{GonBlaMarBul22} A. Gonoskov, T. G. Blackburn, M. Marklund, and S.
S. Bulanov, \emph{Charged particle motion and radiation in strong
electromagnetic fields}, Rev. Mod. Phys.\thinspace \textbf{94}, 045001
(2022).

\bibitem{Fedotov-etal23} A. Fedotov, A. Ilderton, F. Karbstein \textit{%
et.al., \emph{Advances in QED with intense background fields}},\textit{\ }%
Phys. Rep. \textbf{1010}, 1 (2023).

\bibitem{Sarri-etal25} G. Sarri, B. King, T. Blackburn \textit{et.al.}, 
\emph{Input to the European Strategy for Particle Physics: Strong-Field
Quantum Electrodynamics}, arXiv:2504.02608.

\bibitem{LanLi71} L. D. Landau and E. M. Lifshitz, \emph{The classical
theory of fields},\emph{\ }(Pergamon Press, Oxford, 1971).

\bibitem{SokTerBagGalShu68} A. A. Sokolov, I. M. Ternov, V. G. Bagrov, D. V.
Gal'tsov, and V. Ch. Zhukovskii, \emph{\"{U}ber die selbst\"{a}ndige
Polarisierung des Spins der Elektronen durch Ausstrahlung bei der Bewegung
auf einer Spiralbahn im Magnetfeld}, Zeit. Physik. \textbf{211}, 1 (1968).

\bibitem{SokZhuKol69} A. A. Sokolov, V. Ch. Zhukovskii and M. M.
Kolesnikova, Izv. Vyssh. Uchebn. Zaved. Fiz. \textbf{2}, 108 (1969).

\bibitem{SokGalKol71} A. A. Sokolov, D. V. Gal'tsov, and M. M. Kolesnikova,
Izv. Vyssh. Uchebn. Zaved. Fiz. \textbf{4}, 14 (1971).

\bibitem{Umov1874} N. Umow, \emph{Ableitung der Bewegungsgleichungen der
Energie in continuirlichen K\"{o}rpern}, Zeit. Math. Phys. \textbf{19}, 418
(1874).

\bibitem{Poynt1884} H. J. Poynting, \emph{On the Transfer of Energy in the
Electromagnetic Field}, Phil. Trans. R. Soc. \textbf{175}, 343 (1884).

\bibitem{Heavi1884} O. Heaviside, \emph{The Induction of Currents in Cores},
Electrician \textbf{13}, 133 (1884); \emph{Electrical Papers}, Vol. 1
(Macmillan, London 1892).

\bibitem{Strat41} J. A. Stratton, \emph{Electromagnetic Theory}
(McGraw-Hill, New York 1941).

\bibitem{NikRit69} A. I. Nikishov and V. I. Ritus, \emph{Radiation spectrum
of an electron moving in a constant electric field}, Zh. Eksp. Teor. Fiz. 
\textbf{56}, 2035 (1969) [Sov. Phys. JETP \textbf{29}, 1093 (1969)].

\bibitem{Nikishov71} A. I. Nikishov, \emph{Quantum processes in a constant
electric field}, Zh. Eksp. Teor. Fiz. \textbf{59}, 1262 (1970) [Sov. Phys.
JETP \textbf{32}, 690 (1971)].

\bibitem{BagGiSF20} V. G. Bagrov, D. M. Gitman, A. A. Shishmarev and A. J.
D. Farias, \emph{Quantum states of an electromagnetic field interacting with
a classical current and their applications to radiation problems}, J.
Synchrotron Rad. \textbf{27}, 902 (2020).

\bibitem{ShiLeBG21} A. A. Shishmarev, A. D. Levin, V. G. Bagrov, D. M.
Gitman, \emph{Semiclassical Description of Undulator Radiation}, J. Exp.
Theor. Phys. (JETPh) \textbf{132}, 247 (2021).

\bibitem{TGJB} T. C. Adorno, A. I. Breev, A. J. D Farias Jr, and D. M.
Gitman, \emph{Electromagnetic radiation of accelerated charged particle in
the framework of a semiclassical approach}, Annal. Phys. \textbf{7}, 535,
(2023).

\bibitem{BagGit2000} V. G. Bagrov and D. M. Gitman, \emph{The Dirac equation
and its solutions }(Walter de Gruyter GmbH \& Co KG, Berlin/Boston, 2014).

\bibitem{NikRit67} A. I. Nikishov and V. I. Ritus, Zh. Eksp. Teor. Fiz. 
\textbf{52}, 1707 (1967) [Sov. Phys. JETP \textbf{25}, 1135 (1967)].

\bibitem{HeiIld09} T. Heinzl and A. Ilderton, Opt. Commun. \textbf{282},
1879 (2009).

\bibitem{RufVerXue10} R. Ruffini, G. Vereshchagin, and S.-S. Xue, Phys. Rep. 
\textbf{487}, 1 (2010).

\bibitem{Watson} G. N. Watson, \emph{Theory of Bessel Functions} (Cambridge
University Press, Cambridge, 1944).

\bibitem{Watsonb} G. N. Watson, \emph{A Treatise on the Theory of Bessel
Functions}, (Cambridge University Press, 2nd ed., Cambridge, England, 1965).

\bibitem{Sprangle93} E. Esarey, S. K. Ride, and P. Sprangle, \emph{Nonlinear
Thomson scattering of intense laser pulses from beams and plasmas}, Phys.
Rev. E \textbf{48}, 3003 (1993).

\bibitem{Heitl36} W. Heitler, \emph{The Quantum Theory of Radiation},\emph{\ 
}(Oxford Univ. Press, London, 1936).

\bibitem{Sakurai} J. J. Sakurai, \emph{Advanced Quantum Mechanics},
(Addison-Wesley, 1st ed., Menlo Park, California, 1967).

\bibitem{AkhBe65} A. I. Akhiezer and V. B. Berestetskii, \emph{Quantum
Electrodynamics}, (Interscience Publishers, New York, 1965).

\end{thebibliography}
\end{document}